\documentclass[twocolumn]{aastex63}

\usepackage{amsmath}
\usepackage{xcolor}
\definecolor{vincent}{rgb}{0.1, 0.0, 0.7}

\submitjournal{ApJS}

\shorttitle{Stellar Polarization Compilation}
\shortauthors{Panopoulou et al.}

\usepackage{lineno}

\begin{document}

\title{A compilation of optical starlight polarization catalogs}

\correspondingauthor{G. V. Panopoulou}
\email{georgia.panopoulou@chalmers.se}

\author{G. V. Panopoulou}
\affiliation{Department of Space, Earth and Environment, Chalmers University of Technology, 412 93, G\"{o}teborg, Sweden}

\author{L. Markopoulioti}
\affiliation{Department of Physics, University of Crete, Voutes University Campus, GR-70013 Heraklion, Greece}

\author{F. Bouzelou}
\affiliation{Department of Physics, University of Crete, Voutes University Campus, GR-70013 Heraklion, Greece}

\author{M. A. Millar-Blanchaer}
\affiliation{Physics Department, University of California, Santa Barbara, CA 93106, USA}

\author{S. Tinyanont}
\affiliation{National Astronomical Research Institute of Thailand, 260 Moo 4, Donkaew, Maerim, Chiang Mai 50180, Thailand}

\author{D. Blinov}
\affiliation{Department of Physics, University of Crete, Voutes University Campus, GR-70013 Heraklion, Greece}
\affiliation{Institute of Astrophysics, Foundation for Research and Technology - Hellas, Vasilika Vouton, GR-70013 Heraklion, Greece}

\author{V. Pelgrims}
\affiliation{Department of Physics, University of Crete, Voutes University Campus, GR-70013 Heraklion, Greece}
\affiliation{Institute of Astrophysics, Foundation for Research and Technology - Hellas, Vasilika Vouton, GR-70013 Heraklion, Greece}

\author{S. Johnson}
\affiliation{Cahill Center for Astronomy and Astrophysics, California Institute of Technology, Pasadena, CA, 91125, USA}

\author{R. Skalidis}
\affiliation{Cahill Center for Astronomy and Astrophysics, California Institute of Technology, Pasadena, CA, 91125, USA}

\author{A. Soam}
\affiliation{Indian Institute of Astrophysics, II Block, Koramangala, Bengaluru 560034, India}

\begin{abstract}

Polarimetry of stars at optical and near-infrared wavelengths is an invaluable tool for tracing interstellar dust and magnetic fields. Recent studies have demonstrated the power of combining stellar polarimetry with distances from the \textit{Gaia} mission, in order to gain accurate, three-dimensional information on the properties of the interstellar magnetic field and the dust distribution. However, access to optical polarization data is limited, as observations are conducted by different investigators, with different instruments and are made available in many separate publications. To enable a more widespread accessibility of optical polarimetry for studies of the interstellar medium, we compile a new catalog of stellar polarization measurements. The data are gathered from 81 separate publications spanning two decades since the previous, widely-used agglomeration of catalogs by Heiles (2000). The compilation contains a total of 55,742 measurements of stellar polarization. We combine this database with stellar distances based on the \textit{Gaia} Early Data Release 3, thereby providing polarization and distance data for 42,482 unique stars. We provide two separate data products: an Extended Catalog (containing all polarization measurements) and a Unique Source catalog (containing a subset of sources excluding duplicate measurements). We propose the use of a common tabular format for the publication of stellar polarization catalogs to facilitate accessibility and increase discoverability in the future.
\end{abstract}

\keywords{polarization --- 
ISM --- catalogs --- surveys}


\section{Introduction} \label{sec:intro}

The polarization of starlight due to the interstellar medium (ISM) is a powerful probe of the interstellar magnetic field. This interstellar polarization arises as a result of dichroic absorption of the starlight by dust grains which are aligned with the ambient magnetic field \citep{Davis1951,Lazarian2007,Andersson2015}. Ever since its discovery \citep{Hiltner1949,Hall1949}, interstellar polarization of starlight has been invaluable for our understanding of the Galaxy's magnetic field \citep{Ellis1978,Heiles1996} as well as its effects on ISM structure \citep{Heyer2008,Clark2014} and star formation \citep{Myers1991,Alves2008,Li2013,Marchwinski2012,Franco2015ApJ...807....5F,Panopoulou2016,Kandori2017}. Stellar polarimetry has also played a decisive role in determining the properties of ISM dust grains \citep{Coyne1974,Serkowski1975,Whittet1978,Whittet2001,Voshchinnikov2016,Siebenmorgen2017,Andersson2022} and their interaction with the magnetic and radiation fields \citep{Whittet1992,Lazarian1997, Arce1998,Andersson2011} \citep[see also][for a recent review of observational constraints on dust models from multi-wavelength polarimetry]{Hensley2021}.

In particular, the combination of stellar polarization with distance information is a powerful probe of the three-dimensional properties of the magnetic field. Studies have placed constraints on the geometry of the field from just outside the heliosphere \citep{Frisch2015}, to nearby structures like the Local Bubble and radio loops \citep{Leroy1999,Andersson2006,Santos2011ApJ...728..104S, Berdyugin2014AA...561A..24B,Panopoulou2021,Turic2021}, portions of spiral arms \citep{Bijas2022MNRAS.515.3352B} and the large-scale magnetic field of the Galaxy \citep{Pavel2014}, even beyond the Galactic center \citep{Zenko2020}. Stellar distance and polarization have also been used recently to place more stringent constraints on the efficiency of dust grain alignment with the magnetic field \citep{Medan2019ApJ,Panopoulou2019AA...624L...8P,Planck2020,Piccone2022,Angarita2023}.

With the advent of \textit{Gaia} \citep{Gaia_mission}, distances from stellar parallax measurements have become available for billions of stars \citep{Bailer-Jones2021}. The combination of this information with stellar photometric surveys \citep[e.g.][]{Skrutskie2006,Panstarrs2010} has revolutionized our view of the 3D structure of the interstellar dust distribution within a few kiloparsecs from the Sun \citep{Green2019,Leike2019,Lallement2019,Leike2020,Lallement2022}. Similar progress can be made in our understanding of the magnetic field geometry by combining stellar polarimetry with \textit{Gaia} distances in a similar fashion \citep{Panopoulou2019ApJ...872...56P,Pelgrims2023}. However, substantial volumes of stellar polarization measurements are necessary for creating such reconstructions. With most existing catalogs of stellar polarization having samples of hundreds to thousands of stars, the amount of available data is the main limiting factor towards progress on this front.

Existing data on stellar polarization are published in many different catalogs, making their combined use difficult. In particular, the most widely-used database for stellar polarimetry is that presented by \citet{Heiles2000AJ....119..923H}, which contains 9286 stars. Other commonly used catalogs include \citet{Berdyugin2001AA...372..276B,BerdyuginTeerikorpi2001AA...368..635B,Berdyugin2004AA...424..873B,Berdyugin2014AA...561A..24B}, providing data at high galactic latitudes. 

As discussed above, the utility of stellar polarization catalogs increases tremendously with the addition of stellar distance information. For this reason, several recent works have performed cross-matches of stellar polarization catalogs with \textit{Gaia}. \citet{Gontcharov2019} performed a cross-match of 13 polarization catalogs \citep[including][]{Heiles2000AJ....119..923H, Berdyugin2014AA...561A..24B} with \textit{Gaia} Data Release 2
(DR2) for stars out to 500 pc, providing information for 3871 stars.
\citet{Meng2021ChAA..45..162M} cross-matched the \textit{Gaia} DR2 catalog with the starlight polarization catalog from
\citet{Heiles2000AJ....119..923H} to obtain precise distance measurements together
with polarization information for 7613 stars. \citet{Versteeg2023} recently presented previously unpublished data from the Interstellar Polarization Survey, with optical polarization measurements of 40,000 stars - most of which were identified in \textit{Gaia} DR2 (publication of the data is anticipated in the near future). 
The largest survey of stellar polarization has been recently published by \citet{Clemens2020}. The Galactic Plane Infrared Polarization Survey (GPIPS) is the only stellar polarization survey that contains data for millions of sources. The catalog contains polarization measurements in the NIR (H-band) as well as cross-matches to the Two Micron All-Sky Survey \citep[2MASS][]{Skrutskie2006} and \textit{Gaia}. Other notable NIR polarization surveys include: \citet{Nishiyama2009,Sugitani2011,Hatano2013}. 
Millions of optical polarization measurements are anticipated in the near future by planned surveys \citep{SouthPol2012,Tassis2018}.

Our aim is to provide the community with a large database of stellar polarization measurements in the optical combined with \textit{Gaia} distances, tailored for studies of the ISM. We exclude catalogs that target primarily intrinsically polarized sources (young stellar objects, active galactic nuclei, Be-stars etc). We also restrict our compilation to data published after the \citet{Heiles2000AJ....119..923H} agglomeration. We note that the \citet{Heiles2000AJ....119..923H} agglomeration is not complete $-$ there were many publications prior to 2000 that were not included in that database. However, due to the large time commitment necessary to transcribe data from non-machine-readable format to a common catalog, we have restricted our data collection to publications post-2000 (with few exceptions, as noted in Section \ref{sec:data} and Appendix \ref{appendix:catalogs}). The compilation is primarily comprised of optical data, as our goal is the combination of stellar polarization with \textit{Gaia}. Even though the GPIPS survey does provide cross-matches with \textit{Gaia}, we do not include it in our catalog as the data volume is so large (and data discoverability for this large survey is not an issue). The compilation we present here is complementary to GPIPS, spanning a much larger sky area, but mostly restricted to stars within a few kpc of the Sun (Fig. \ref{fig:G-dist}). Finally, we do not include catalogs of spectro-polarimetry \citep[unless measurements in standard broad band filters, obtained with convolution of their bandpasses with the spectra are provided, e.g.][]{Bagnulo2017AA...608A.146B}, restricting our catalog to have a simpler format than would be required for publishing full spectra. 

The data that are included in our compilation represent a decades-long community effort to improve our knowledge of interstellar magnetic fields. The data span a wide range of environments, from the diffuse ISM at high galactic latitudes \citep{Berdyugin2014AA...561A..24B}, to molecular clouds \citep[e.g.][]{Pereyra2004ApJ...603..584P,Wang2017ApJ...849..157W}, HII regions \citep{Santos2012ApJ...751..138S}, stellar clusters \citep[e.g.][]{Eswaraiah2011MNRAS.411.1418E, Topasna2018AA...615A.166T}, the Large Magellanic Cloud \citep{Lobo2015ApJ...806...94L} and more. 

We discuss our selection of stellar polarization catalogs in Section \ref{sec:collection}. As the data from the different catalogs are very inhomogeneous, special treatment was necessary for incorporating all data into a common tabular format (Sections \ref{sec:homogenization}, \ref{sec:homogenization_p}, \ref{sec:homogenization_meta}). Notes for the treatment of specific catalogs can be found in Appendix \ref{appendix:catalogs}. The process of cross-matching the polarization compilation with \textit{Gaia} and obtaining stellar distances is described in Section \ref{sec:gaia}, along with certain considerations. We describe the data products of our compilation in Section \ref{sec:products}. We present properties of the compilation in Section \ref{sec:results}. We propose a common standard for the publication of future stellar polarization catalogs to enable continued access and discoverability in Section \ref{sec:future}.

\section{Archival Data}
\label{sec:data}

\subsection{Selection of stellar polarization catalogs}
\label{sec:collection}

\begin{table*}[]
    \centering
    \begin{tabular}{c|c|c||c|c|c}
        RefID & Publication &  FilterID(s) & RefID & Publication &  FilterID(s) \\
        \hline
        0 & \citet{Vrba1976AJ.....81..958V} &  0 & 40 &\citet{Prokopjeva2014JQSRT.146..410P}& 20 \\
        1 & \citet{Moneti1984ApJ...282..508M} &  0, 12, 13,17 & 41 & \citet{Chakraborty2014MNRAS.442..479C}&  13 \\
        2 & \citet{Heyer1987ApJ...321..855H}&  0 & 42 & \citet{Berdyugin2014AA...561A..24B}&  0, 13 \\
        3 & \citet{Goodman1990ApJ...359..363G}&  30 & 43 & \citet{Alves2014AA...569L...1A} & 20\\
        4 & \citet{Heiles2000AJ....119..923H}&  0 & 44 &  \citet{Soam2015AA...573A..34S} &  13,20,29\\
        5 & \citet{Sen2000AAS..141..175S}  &  0 & 45 & \citet{Panopoulou2015MNRAS.452..715P}&  20\\
        6 & \citet{Whittet2001ApJ...547..872W} &  2,7,13,20,34& 46 &\citet{Lobo2015ApJ...806...94L} &  13\\
        7 & \citet{Oudmaijer2001AA...379..564O}&  2,7,13,26,36 & 47 & \citet{Franco2015ApJ...807....5F} &  20\\
        8 & \citet{BerdyuginTeerikorpi2001AA...368..635B}& 7,13,26 & 48 & \citet{Seron2016MNRAS.462.2266S} &  13, 20, 24\\
        9 & \citet{Berdyugin2001AA...372..276B}&  0  & 49 & \citet{Das2016ApSS.361..381D} &  23\\ 
        10 & \citet{Pereyra2002ApJS..141..469P}&  13 & 50 & \citet{Neha2016AA...588A..45N} &  29\\
        11 & \citet{Menard2002AA...396L..35M} &  32 & 51 & \citet{Cotton2016MNRAS.455.1607C} &  9\\
        12 & \citet{Clarke2002AA...383..580C}&  7,13 & 52 & \citet{Chakraborty2016ApSS.361..321C} &   {20}\\
        13 & \citet{Berdyugin2002AA...384.1050B}&  0 & 53 & \citet{Zejmo2017MNRAS.464.1294Z} &   {20}\\
        14 & \citet{Gil-Hutton2003MNRAS.345...97G}&   {13} & 54 & \citet{Wang2017ApJ...849..157W} &   {23,33}\\
        15 & \citet{Pereyra2004ApJ...603..584P}&   {13} & 55 & \citet{Soam2017MNRAS.465..559S} &   {29}\\
        16 & \citet{Berdyugin2004AA...424..873B}&  0 & 56 & \citet{Soam2017MNRAS.464.2403S} &   {23} \\
        17 & \citet{Poidevin2006ApJ...650..945P}&   {31} & 57 & \citet{Reig2017AA...598A..16R} &   {20}\\
        18 & \citet{Alves2006} &   {7} & 58 &\citet{Cotton2017MNRAS.467..873C} &   {9}\\
        19 & \citet{Fossati2007ASPC..364..503F}&   {6,15,22,32}& 59 & \citet{Bagnulo2017AA...608A.146B} &  {6, 15, 22, 32}\\ 
        20 & \citet{Alves2007AA...470..597A}&   {7}  & 60 & \citet{Topasna2018AA...615A.166T}&   {2,7,13,20,34} \\
        21 & \citet{Andersson2007ApJ...665..369A}&   {2,7,13,20,34} & 61 & \citet{Soam2018MNRAS.476.4782S} &   {29} \\
        22 & \citet{Weitenbeck2008AcA....58...41W}&  0,  {2,7,13,20,34}& 62 &\citet{Slowikowska2018MNRAS.479.5312S}&   {20}\\
        23 & \citet{Weitenbeck2008AcA....58..433W}&  0,  {2,7,13,20,34} & 63 & \citet{Neha2018MNRAS.476.4442N} &   {13,29,34}\\
        24 & \citet{Medhi2008MNRAS.388..105M}&   {7,13,23} & 64 & \citet{Hutsem2018AA...620A..68H} &   {16}\\
        25 & \citet{Medhi2010MNRAS.403.1577M}&   {7,13,23}  & 65 & \citet{Panopoulou2019ApJ...872...56P} &   {20}\\
        26 & \citet{Bailey2010MNRAS.405.2570B}&   {27}& 66 & \citet{Panopoulou2019AA...624L...8P} &   {20}\\
        27 & \citet{Andersson2010ApJ...720.1045A}&   {2,7,13,20,34} & 67 & \citet{Eswaraiah2019ApJ...875...64E} &   {20}\\
        28 & \citet{Targon2011ApJ...743...54T}&   {20}  & 68 & \citet{Cotton2019MNRAS.483.3636C} &   {5, 9, 19, 24}\\
        29 & \citet{Santos2011ApJ...728..104S}&   {13} & 69& \citet{Choudhury2019MNRAS.487..475C} &   {20} \\
        30 & \citet{Eswaraiah2011MNRAS.411.1418E}&  { 7, 13, 23, 34} & 70 & \citet{Vaillancourt2020ApJ...905..157V}  &  {2, 7, 13, 26, 36}\\
        31 & \citet{Alves2011AJ....142...33A}&   {20}& 71 & \citet{Topasna2020PASP..132d4301T} &   {7, 13, 20, 34}\\
        32 & \citet{Santos2012ApJ...751..138S}&  0,  {13, 20,34} & 72 & \citet{Singh2020AJ....160..256S} &   {7, 13, 26, 34} \\
        33 & \citet{Eswaraiah2012MNRAS.419.2587E}&   {7, 13, 23, 34} & 73 & \citet{Singh2020AJ....159...99S} &   {7, 13, 26, 34}\\
        34 & \citet{Soam2013MNRAS.432.1502S}&   {20}  & 74 & \citet{Piirola2020AA...635A..46P}&   {10}\\
        35 & \citet{Pandey2013ApJ...764..172P}&  0,  {13, 20}& 75& \citet{Topasna2021PASP..133j4301T} &   {2,7,13,20,34}\\
        36 & \citet{Eswaraiah2013AA...556A..65E}&   {7, 13, 23, 34} & 76 &\citet{Uppal2022AJ....164...31U}&  {33}\\
        37 & \citet{Andersson2013ApJ...775...84A}&   {full list}& 77 &\citet{Topasna2022PASP..134h4301T}&  {7, 13, 20, 34} \\
           &                   &  {in footnote}\footnote{1,  2,  3,  4,  7,  8, 11, 13, 14, 18, 21, 25, 26, 28, 35, 36, 37, 38, 40, 41, 42, 43} & 78 & \citet{Singh2022MNRAS.513.4899S} &   {7, 13, 20, 34}\\
        38 & \citet{Santos2014ApJ...783....1S}&   {13, 20, 34} & 79 & \citet{Choudhury2022RAA....22g5003C} &   {20}\\
        39 & \citet{Reig2014MNRAS.445.4235R} &   {20} & 80 &\citet{Bijas2022MNRAS.515.3352B}&   {39}\\
    \end{tabular}
    \caption{Publications from which we have included data in our compilation. Columns specify: the reference identifier (\texttt{RefID}), the bibliographic reference to the publication, and the FilterID used to specify the observing filter for each observation ( {see T}able \ref{tab:filters}). }
    \label{tab:references}
\end{table*}

Data on stellar polarization have been presented in many different publications, compiled by different investigators and with different formats. To construct a unified database of optical polarization for ISM studies, our first step was to search the literature for published catalogs. We adopted the following search strategies. We used the SAO/NASA Astrophysics Data System (ADS) system to search for publications containing the words ``optical" and ``polarization" in their abstract, as well as the keyword ``polarimetry" or the keyword: ``starlight polarization". We inspected hundreds of publications, retaining only those with the following criteria:
\begin{itemize}
    \item Publication date post-2000.
    \item Public catalog of stellar polarization in the optical.
    \item Data with polarization arising primarily due to the ISM.
\end{itemize}
We also searched for publications citing the \citet{Heiles2000AJ....119..923H} agglomeration. Finally, we queried the Vizier\footnote{DOI : 10.26093/cds/vizier} system for catalogs containing the identifiers ``optical" and ``polarization". This final search returned very limited results compared to our aforementioned wider search. We note that $\sim 75\% $ of the catalogs included in our final compilation did not appear in the Vizier search. 

We further limited our selection of catalogs to those making polarization data of individual stars publicly available. Some catalogs only contained polarization properties averaged over a field of view. These were not included in our agglomeration. We also restricted our compilation to catalogs providing adequate information to obtain relative linear Stokes parameters ($q,u$) of individual stars.  {In some instances, data were provided after subtraction of a foreground ISM component \citep[e.g.][]{Topasna2020PASP..132d4301T}. We did not include such foreground-subtracted data, opting instead to only have the original measurements prior to any such post-processing}. Some catalogs did not publish stellar coordinates or names with which to identify stars. If stellar membership identifiers were provided for stars towards open clusters, in some cases we were able to obtain stellar coordinates from the WEBDA\footnote{\url{https://webda.physics.muni.cz/webda.html}} database.

We have made an exception to the date constraint by including data from \citet{Vrba1976AJ.....81..958V, Moneti1984ApJ...282..508M, Heyer1987ApJ...321..855H, Goodman1990ApJ...359..363G}. These data were part of the \citet{Heiles2000AJ....119..923H} agglomeration. However, the uncertainties of the polarization measurements for these stars were not provided in the latter catalog, and only approximate coordinates were given. By using the data as originally published we have been able to provide accurate \textit{Gaia} matches to most of these stars. More details can be found in  Appendix \ref{appendix:catalogs}.

A list of the 81 publications comprising our final data compilation is presented in Table \ref{tab:references}. The Table presents the reference identifier (\texttt{RefID}) used to specify the publication within our catalog, a bibliographic link to each publication, as well as the observing bands used (specific information on the filters is given in Table \ref{tab:filters}).

The catalogs presented in Table \ref{tab:references} are very inhomogeneous. The data are provided in different formats. The available information varies from catalog to catalog. Some catalogs provide relative linear Stokes parameters.  {S}ome provide fractional linear polarization, $p$, and  {the Electric Vector Position Angle - also often referred to as the polarization angle - PA}.  {S}ome do not provide measurement uncertainties, some do not provide stellar coordinates, measurements are done at different wavelengths (some times multiple measurements of the same star are provided, e.g. in different filters). Some catalogs provided data in machine-readable format and some did not. For the latter we have transcribed data by hand or by copying tables from the .html/.pdf version of the manuscript.

Compiling one common, homogeneous catalog of stellar polarization from these different catalogs required substantial effort and specific treatment. We describe the details of our treatment of certain catalogs in Appendix \ref{appendix:catalogs}. In the following, we describe the kind of data we retain for our final compilation.

\subsection{Homogenization of stellar coordinates and stellar identifiers}
\label{sec:homogenization}
We aimed to obtain stellar coordinates in the International Celestial Reference System (ICRS) equivalent to the Fifth Fundamental Catalog (FK5) J2000. We convert all Right Ascension (RA) and Declination (Dec) values to decimal format, in units of degrees. 

When coordinates were provided in B1950 \citep{Vrba1976AJ.....81..958V,Heyer1987ApJ...321..855H,Goodman1990ApJ...359..363G,Alves2006,Targon2011ApJ...743...54T}, we transformed these to J2000 taking into account only precession and not proper motions. The effect of proper motions will later be taken into account when performing the cross-match with \textit{Gaia} Early Data Release 3 \citep[EDR3][]{GaiaEDR3}, at which stage proper-motion information becomes available (Section \ref{sec:gaia}). 

If coordinates are provided for epochs later than J2000 we transform those to J2000 (see discussion on individual catalogs in Appendix \ref{appendix:catalogs}). 
Some catalogs did not provide coordinates, but did provide some means of stellar identification. If the names of stars were provided in standard catalogs such as Henry Draper (HD) and Bonner Durchmusterung (BD) \citep{Cannon1918,Argelander1903}, we queried Simbad via the function
\begin{verbatim}
astroquery.Simbad.query_object
\end{verbatim} \citep{Ginsburg2019AJ....157...98G}  to obtain the coordinates in the celestial reference frame. For some stars in the \citet{Berdyugin2001AA...372..276B} catalog, the names given referred to catalogs that could not be searched (e.g. unpublished data). These entries were excluded from the combined catalog. If stellar identifications were based on cluster membership catalogs provided in the WEBDA\footnote{\url{https://webda.physics.muni.cz/}} service, we used that service to obtain stellar coordinates \citep[e.g. for][]{Topasna2018AA...615A.166T, Topasna2020PASP..132d4301T, Topasna2021PASP..133j4301T, Topasna2022PASP..134h4301T}. Some catalogs provide other names for their targets, which can be used for stellar identification. For example, they may provide the source identification number from a \textit{Gaia} data release. In these cases we have queried the corresponding \textit{Gaia} database to obtain coordinates.

Obtaining accurate coordinates for stars in the \citet{Heiles2000AJ....119..923H} catalog required special treatment and is described in detail in Appendix \ref{appendix:catalogs}. Despite our best efforts, sometimes a stellar identification is uncertain or unobtainable. In these cases, stellar coordinates are assigned to `Not A Number' (NAN).

We assign each star in a given catalog a unique identification number, \texttt{starID}. We also retain two columns for stellar identification information as provided in the original catalog: a Name column and a \texttt{rawstarID} column. The former is a string, and can often contain an HD number. For catalogs that published a \textit{Gaia} identification number, we set the `Name' column to \texttt{GaiaDRX\_},where X can be 1, 2 or 3, followed by the Gaia identifier from DR1, DR2, etc. The \texttt{rawstarID} column often contains a numerical identifier (e.g. the row of the original table that the star was in). These can be used to track a given measurement in our final compilation to its original catalog. If multiple measurements of the same star are given in a catalog, with the same \texttt{rawstarID}, these also have the same \texttt{starID} in our final catalog. More information on the columns contained in our final compilation can be found in Section {\ref{sec:products}.}

\subsection{Homogenization of polarization measurements}
\label{sec:homogenization_p}
 {For each source we compile} measurements of  {the fractional linear polarization} $p$ and its uncertainty $e_p$ (expressed in fraction, not percentage),  {the polarization angle PA and its uncertainty $e_{\rm PA}$.}  {We also obtain} relative linear Stokes parameters $q,u$ and their uncertainties $e_q$, $e_u$  {(in fraction, not percentage)}.  {The Stokes parameters $q = Q/I$ and $u = U/I$ have been normalized to the total intensity I, as is common convention in stellar polarimetry studies}. All uncertainties are $1\sigma$.

In our final polarization catalog, each line corresponds to a single measurement. Therefore, measurements of the same star made using e.g. different filters or observed at different times are presented in multiple lines. 

We work with the polarization fraction $p$, which is the biased estimator for the true linear polarization fraction in the low signal-to-noise regime. If catalogs provide a de-biased estimator, we convert the data back to biased $p$ according to the prescription adopted by each publication. We convert all $p$ and $e_p$ to fractions, defined within the range [0, 1]. 

 {We use the PA in the equatorial reference frame, whenever such information is given.} We convert all  {PA} and their uncertainty to degrees within the range  {[0$^\circ$, 180$^\circ$]}. To avoid confusion, we distinguish between the polarization angle expressed in degrees  {(PA)} and that in radians ($\theta$, and its uncertainty $e_\theta$).  {In the cases when PA were not reported, the studies reported Stokes parameters (also in the equatorial reference frame). } If Stokes parameters are provided, we calculate the following quantities:
\begin{equation}
\begin{split}
    p &= \sqrt{q^{2} + u^{2}},\hspace{2mm} e_{p} = \sqrt{\frac{q^{2}
	e^{2}_{q}+u^{2}e^{2}_{u}}{q^{2}+u^{2}}}\\
    PA &= 0.5 \arctan(u/q) \cdot 180^\circ /\pi,
\end{split}
\label{eqn:evpa}
\end{equation}
where the 2-argument arctangent is used. 

 {For all catalogs that provided $p$ and PA, we calculate the Stokes parameters via}:
\begin{equation}
\begin{split}
    q &= p \, \cos(2 \, \theta)\\
    u &= p \, \sin(2 \, \theta) \\
    e_q&=\sqrt{(e_p \cos(2 \, \theta)^2+(2\, u \, e_{\theta})^2}\\
    e_u&=\sqrt{(e_p \sin(2 \, \theta)^2+(2 \, q\, e_{\theta})^2}
\end{split}
\label{eqn:querrs}
\end{equation}
We calculate the debiased polarization fraction $p_d$ as \citep{Plaszczynski2014}:
\begin{equation}
p_d = p - e_p^2 \frac{1-e^{-p^2/e_p^2}}{2p}.
\label{eq:pd}
\end{equation}

 {A small number of studies only published Stokes parameters (not $p$, PA) \citep{Clarke2002AA...383..580C,Gil-Hutton2003MNRAS.345...97G,Cotton2016MNRAS.455.1607C,Zejmo2017MNRAS.464.1294Z,Slowikowska2018MNRAS.479.5312S}. In these cases we computed $p$, PA from equations \ref{eqn:evpa} and verified that the Stokes parameters had been provided after calibration. For the data from \citet{Vaillancourt2020ApJ...905..157V} we used the published Stokes parameters, as the provided polarization fractions were debiased (hence we could not recover the biased $p$ for sources where $p_d$ was set to 0). For \citet{Weitenbeck2008AcA....58...41W} and \citet{Weitenbeck2008AcA....58..433W} we calculated $p$ and PA from the Stokes parameters and associated uncertainties, as the systematic errors needed to be added separately (see appendix \ref{appendix:catalogs}). In the instances where all information was provided (Stokes parameters, polarization fractions, {PAs}) we checked for consistency between the p, {{PA} computed from the Stokes parameters and those given in the catalogs. We found only two instances where some measurements were inconsistent (see appendices \ref{app:weitenbeck+}, \ref{app:piirola}).}

 {We calculate t}he uncertainty on the polarization angle by evaluating the integral:
\citet{Naghizadeh1993}: 
\begin{equation}
 \int_{-1e_{\theta}}^{1e_{\theta}} G(\theta;P_0) d(\theta) = 68.27\%,
 \label{eqn:sigmatheta}
\end{equation} where $P_0 = p_d/e_p$.
In equation \ref{eqn:sigmatheta}, $G$ is the probability density function defined as:
\begin{equation}
\begin{split}
 G(\theta;\theta_0;P_0) &= \\
  \frac{e^{-\frac{P_0^2}{2}}}{\sqrt{\pi}} &\left\{ \frac{1}{\sqrt{\pi}} + 
  \eta_0 e^{\eta_0^2}
 [ 1 + erf(\eta_0) ] \right\},
\end{split}
\end{equation}
where $\eta_0 = P_0/\sqrt{2}\cos{2(\theta - \theta_0)}$ and $erf$ is the 
Gaussian error function.  {In the special case where neither the Stokes parameters nor the uncertainty in  {PA} are provided, we first calculate the `biased' estimate of the polarization angle uncertainty: $ \hat{e}_\theta = 0.5 \, e_p/p$. This is used to calculate the Stokes parameter uncertainties via equations \ref{eqn:querrs} - as this inverts the process of how the original $p, \theta$ were computed from the observed Stokes parameters. The final reported uncertainty on the  {PA} in our catalog comes from equation \ref{eqn:sigmatheta}}.

If not enough information is given to obtain any of the above, (most commonly the $e_p$ and $e_{PA}$) we set the corresponding values to NAN. In some cases, the original catalog entries had $e_p = 0$. This was the case for 124 lines in \citet{Heiles2000AJ....119..923H}, and for 1 measurement in each of the following catalogs: \citet{Berdyugin2001AA...372..276B, Berdyugin2014AA...561A..24B, Soam2017MNRAS.464.2403S, Neha2018MNRAS.476.4442N}. This likely is a typographical error (e.g. not enough decimal places were provided in the catalog entries for the polarization fraction uncertainty). We therefore set $e_p$ to NAN for all entries with $e_p = 0$ in their original catalogs.

In cases when systematic uncertainties are not included in the catalog, but are quoted in the text, we add these in quadrature to the statistical uncertainty \citep[e.g.][]{Weitenbeck2008AcA....58..433W}.

The  {PA} is defined with respect to the North Celestial Pole and increasing to the East in the reference frame of the observations. Because different reference frames are used in the various catalogs in our compilation, this means that there are systematic differences in the  {PA} due to this effect. Most of the observations in the catalogs we consider have been conducted after the year 2000. We assume that these  {PA} measurements were tied to the ICRS or FK5 (J2000) reference frames, thus no correction is necessary. This is our motivation for choosing to report coordinates in epoch 2000 so as to be consistent with the  {PA} reference point for the majority of the catalogs. 

However, the catalogs of \citet{Heiles2000AJ....119..923H,Vrba1976AJ.....81..958V,Moneti1984ApJ...282..508M,Heyer1987ApJ...321..855H,Goodman1990ApJ...359..363G} contain measurements from earlier epochs. In particular, \citet{Heiles2000AJ....119..923H} includes measurements tied to many different reference frames, some dating back to B1900 \citep[e.g.][]{Hall1958PUSNO..17..275H,Appenzeller1968}. To the best of our knowledge, the  {PA} from the original catalogs that were incorporated into \citet{Heiles2000AJ....119..923H} and earlier \citet{Ellis1978, Mathewson1970} have not been corrected for the rotation of the reference frame (we found no mention of such a correction in those works). 

Identifying the epoch to which an individual measurement was performed is complicated for the \citet{Heiles2000AJ....119..923H} dataset. First, the majority of measurements have been reported from a previous compilation, also comprised of earlier compilations, making tracing the original source of the observations difficult. Second, duplicate entries (multiple measurements of the same star from different catalogs) have been averaged in \citet{Heiles2000AJ....119..923H}, precluding a correction. 

We choose to perform no correction for the coordinate reference frame change on the  {PA}, primarily because the effect is smaller than the quoted measurement uncertainties for the majority of sources in the aforementioned catalogs. The rotation would be most severe for stars near the celestial poles, of which there are few in the aforementioned catalogs. For example, a star at declination 80$^\circ$ will show a rotation of the position angle of 1.5 degrees from 1900 to 2000. There are 22 stars in the \citet{Heiles2000AJ....119..923H} catalog with $|b| > 80^\circ$ and none in the \cite{Vrba1976AJ.....81..958V,Moneti1984ApJ...282..508M,Heyer1987ApJ...321..855H,Goodman1990ApJ...359..363G} catalogs. Stars with high proper motion should also be corrected for rotation of the  {PA}, however in \citet{Heiles2000AJ....119..923H} the maximum star proper motion is $\sim 500$ mas, which corresponds to a minor change of the  {PA}. Since this effect does not significantly alter the  {PA} for the majority of stars in our catalog, we have chosen not to correct any measurements for the rotation of the reference frame. We do caution users of the catalog that systematic rotations may be important for the  {PA} of stars at $|b| > 80 ^\circ$ from \citet{Heiles2000AJ....119..923H}.

\subsection{Homogenization of metadata}
\label{sec:homogenization_meta}
We collect metadata which may be of interest to future users. We homogenize observing filter information by creating a unified filter index with  {52} distinct entries. These are filters with substantially different effective wavelengths $\lambda_{\rm eff}$ that cover all of the bands used in the publications in our compilation. The filter index is shown in Table \ref{tab:filters}. Each polarization measurement is assigned a filter identifier (FilterID) that corresponds to one of the filters in Table \ref{tab:filters}. We have attempted to match the effective wavelengths of the filters in each publication to those of the index to within 2 decimal points (with $\lambda_{\rm eff}$ measured in microns). If an effective wavelength is not quoted in the text of a publication, we searched for previous or later publications from the same authors or instruments to obtain the corresponding $\lambda_{\rm eff}$. If no filter is specified, or the filter $\lambda_{\rm eff}$ is unclear, or in the case of \citet{Heiles2000AJ....119..923H} measurements could have been averaged across many filters, we assign \texttt{FilterID} = 0.

We include the Julian date of the observations. We set this to be the earliest date quoted in the text for the presented set of observations, unless individual observing dates are explicitly stated for each source.

If the publication text mentions a source as having intrinsic polarization (i.e. any contribution to the observed polarization that is not induced by the ISM) that source is flagged. This flag is in the \texttt{IntrPol} column (1 = intrinsically polarized, 0 = not intrinsically polarized or unknown). Intrinsic polarization has been identified by various means in the different publications, for example by the presence of a circumstellar disc, or by detecting time variability in the source polarization.


\begin{table*}[]
    \centering
    \begin{tabular}{c|c|c||c|c|c}
FilterID & Band & $\lambda_{\rm eff} (\mu m)$ & FilterID & Band & $\lambda_{\rm eff} (\mu m)$\\
\hline
0	& No filter or unclear	&	$-$ & 21	& custom & 0.65\\
1	& custom & 0.35 & 22	& R Bessel (FORS1)$\rm^a$	&0.657\\
2	& U	Johnson	& 0.37 & 23	& Rc Johnson-Cousins/AIMPOL &0.67\\
3	& custom\footnote{Refers to the broad-band averaging of spectropolarimetric data in \citet{Andersson2013ApJ...775...84A}.}  & 0.375 & 24	& 650LP HIPPI &0.7\\
4	& custom 		& 0.4 & 25	& custom    &0.7\\
5	& 425SP		& 0.4 & 26	& R Johnson	&0.7\\
6	& B	Bessel (FORS1)\footnote{\url{www.eso.org/sci/facilities/paranal/instruments/fors/doc/VLT-MAN-ESO-13100-1543_v82.pdf	}}	&0.429 & 27	& broad red band	&0.735-0.804\\
7	& B	Johnson		& 0.44 & 28	& custom		&0.75\\
8	& custom	&0.45 & 29	& Rkc	Kron-Cousins/AIMPOL	&	0.76\\
9	& g'	SDSS &0.475 & 30	& RG-645 Schott		&0.763\\
10	& BVR	Dipol2\footnote{weighted-mean} & $-$ & 31	& RG-645	&	0.766\\
11	& custom	&0.5 & 32	& I Bessel	&0.768\\
12	& G  	&0.51 & 33	& i' SDSS	&0.77\\
13	& V 	Johnson &	0.55 & 34	& I Cousins &0.79\\
14	& custom		&0.55 & 35	& custom	 		&0.8\\
15	& V Bessel (FORS1)$\rm^a$		& 0.554 & 36	& I	&0.83\\
16	& V Bessel high-114$\rm^a$		&0.561 & 37	& custom 			&0.85\\
17	& VRI	& 0.5-0.85	 & 38	& custom	 		&0.9\\
18	& custom		&0.6 & 39	& I Johnson	&0.9\\
19	& r'	SDSS		&0.62 & 40	& custom	 		&0.95\\
20	& Rc Cousins &	0.64  & 41 & custom 		&	0.975\\
& & & 42	& custom 		&1\\
& & & 43	& H 		&1.65\\
\hline
    \end{tabular}
    \caption{Filter index. Each polarization measurement is assigned an integer FilterID, corresponding to the filter used during observations (first column). The filter name and effective wavelength are shown in the second and third columns.}
    \label{tab:filters}
\end{table*}

\subsection{Cross-matching with Gaia to obtain accurate stellar coordinates and distances}
\label{sec:gaia}

\textit{Gaia} has obtained parallax measurements to billions of stars, however distances are not directly provided.  {We choose to cross-match our homogenized catalog with \textit{Gaia} EDR3 \citep{GaiaEDR3}, for which \citet{Bailer-Jones2021} have provided probabilistic distance estimates. We explain the process for the cross-match below.}  

 {We use the stellar coordinates obtained from the original polarization catalogs and converted to J2000 (Section \ref{sec:homogenization}). These coordinates are only approximate, as the effect of proper motions has not been taken into account. }
Because stellar coordinates from the original catalogs span a range of epochs, while proper motions are not known for the sources prior to cross-match with \textit{Gaia}, a simple cross-match within a radius of $\sim 2 \arcsec$ could yield miss-identifications or no identifications for sources with proper motions of 30 mas/yr or larger (if B1950 coordinates were given). We therefore use a two-step approach to cross-matching with \textit{Gaia} EDR3. 

We first cross-match our  {homogenized} catalog with the \textit{Gaia} EDR3 catalog with a radius of 14\arcsec. To define  {this} search radius, we take into account the distribution of proper motions $\mu$ of stars in \textit{Gaia} DR2, provided in tables 10.1 and 10.2 of the \textit{Gaia} DR2 documentation\footnote{\url{https://gea.esac.esa.int/archive/documentation/GDR2/Catalogue_consolidation/chap_cu9val_cu9val/sec_cu9val_942/ssec_cu9val_943_pm.html}}. 
The magnitude range with the largest standard deviation in proper motion is that of $G-$band magnitude 8, with a standard deviation of $\sigma_\mu = 106$ mas/yr. 
We thus select a very broad search radius of 14\arcsec, corresponding to the path of a star with proper motion equal to twice the aforementioned standard deviation $\sigma_\mu$ over the period 1950 - 2016. The second step is to query the resulting catalog after the first cross-match within a smaller region of 2\arcsec radius, but now propagating the coordinates of all \textit{Gaia} sources in this region from 2016 to the year 2000 (taking into account both precession and proper-motion information when it is available). 

With this two-step process we obtain 42,482 unique sources with \textit{Gaia} identifications. If we instead compute the cross-match only with a 2\arcsec radius without propagating the Gaia (2016) coordinates to J2000 we obtain 42,143 unique sources. 

If multiple sources are found within the 2\arcsec search radius of a given star in our catalog, we assign the match to be the brightest star among the potential matches. This is motivated by the fact that most of the polarization catalogs have a limiting magnitude much brighter than that of \textit{Gaia} EDR3.  {We note that the brightest star may not be the optimal choice for all sources, however this likely is the case for a very small fraction of the catalog sources.}
Only 1708 stars have multiple matches in the search radius.

 {The \textit{Gaia} sources obtained via this cross-match procedure are compiled with the polarization data in a single catalog (Extended catalog - see Section \ref{sec:products}).}
Since the distance catalog we use for most sources is tied to EDR3, we provide the EDR3 coordinates at epoch J2016{, in addition to J2000 coordinates obtained via epoch-propagation from J2016}. 

Two distance estimates are provided in \citet{Bailer-Jones2021} and we include both in our catalog. The first is the geometric distance estimate, which is a probabilistic measure of the stellar distance based only on parallax measurements. The second is a photogeometric estimate, which also incorporates the photometry (brightness, color) of the star, in addition to parallax information. Their catalog provides the 16-, 50- and 84- percentiles of the posterior distribution of the stellar distance, labeled \texttt{r\_lo\_photogeo, r\_med\_photogeo, r\_hi\_photogeo} for photogeometric and \texttt{r\_lo\_photogeo, r\_med\_photogeo, r\_hi\_photogeo} for geometric distance, respectively. The difference between the median \texttt{r\_med\_photogeo} and each of the other quantiles is the equivalent of a 1$\sigma$ uncertainty, and similarly for the geometric distance estimate.  \citet{Anders2022} also provide distances taking into account other photometric surveys, but their catalog only contains for a subset of sources in EDR3. We note that the \citet{Bailer-Jones2021} distances are based on Galactic stellar distribution priors, and hence do not work well for stars in e.g. the Magellanic clouds. This adversely affects only one of the stellar polarization catalogs in our compilation \citep{Lobo2015ApJ...806...94L} and we urge readers to use caution when using distance estimates of stars in this catalog.

We caution that mis-identifications can occur, especially since the brightness of the targets in each catalog and the accuracy of the given stellar coordinates are unknown. Mis-identifications are more likely for faint ($G > 16$ mag) sources and for sources in crowded fields. 


\begin{table*}[]
\centering
\begin{tabular}{|c|c|c|c|c|}
\hline
Polarimeter & Telescope & Observatory & Polarimeter & RefIDs \\
Name & & & Reference & \\
\hline
AIMPOL & 1.04 m & ARIES & \citet{Rautela2004BASI...32..159R} & 24, 25, 30, 33, 34, 35, 36,\\
& & & &   44, 49, 50, 54, 55, 56,  61, 63,\\
& & & &   67, 68,  69, 72, 73, 78, 79, 80\\ 
\hline
IAGPOL & 0.9 m & CTIO & & 32,   38, 43, 48\\
& 1.5m & CTIO & & 10, 46\\
& 1.6m &  OPD/LNA & \citet{Magalhaes1996}& 29, 31, 38, 47, 48\\
& 0.6m & IAG& & 15, 18, 20, 28, 29, 38, 47, 48\\
\hline
Turpol & 2.6 m NOT & ORM & \citet{Piirola1988prco.book..735P}& 7, 8, 9, 13, 16, 37, 70\\
& 2.15 m & CASLEO & & 16\\
 & 60 cm KVA & ORM & & 8, 13\\
 & 0.6 m, 1.25 m & Tuorla, CAO & & 9\\
\hline
IMPOL & 2m & Girawali IUCAA & \citet{Ramaprakash1998} & 36, 40, 41, 52\\
      & 1.2 m & Mount Abu & & 5\\
\hline
HIPPO & 1.9 m & SAAO& \citet{Potter2010}& 27\\
\hline
HIPPI & 3.9 m AAT & Siding Spring & \citet{Bailey2015} & 51, 58, 68\\
\hline
RoboPol & 1.3 m  & Skinakas & \citet{Ramaprakash2019MNRAS}& 39, 45, 53, 57, 62, 65, 66\\
\hline
DiPol & 60 cm KVA & ORM &\citet{Piirola2005}& 16, 42\\
\hline
DiPol-2 & NOT & ORM & \citet{Piirola2014}& 74 \\
&  60 cm KVA & ORM & &42\\
& 4.2 m WHT & ORM & & 74\\
& 2.2 m UH88 & Maunakea & & 74\\
& 60 cm (T60)& Haleakla & & 74 \\
& 1.27 m (H127)& University of Tasmania & & 74 \\
\hline
HPOL & 0.9 m & Pine Bluff& \citet{Wolff1996}, & 22, 23\\
& 3.5 m WIYN & Kitt Peak & \citet{Nordsieck1996} & 22\\
\hline
FORS 1, 2 & VLT & Paranal & \citet{Patat2006}& 11, 19, 59, 64\\
\hline
PlanetPol & 4.2 m WHT & ORM &\citet{Hough2006}& 26\\
\hline
VMI & 0.5 m & VMI  & \citet{Topasna2013} & 60, 71, 75, 77\\
\hline
TRIPOL & 1m & Lulin & \citet{Sato2019}& 54 \\
\hline
EMPOL & 1.2 m & Mount Abu & \citet{Ganesh2020} & 76\\
\hline
ALFOSC & 2.6 m NOT & ORM & (1) & 37\\
\hline
UCTP& 1.9m & SAAO &\citet{Cropper1985}& 21\\
\hline
Beauty \& the Beast & 1.6m & OMM & \citet{Manset1995} & 17\\
\hline
CASPROF & 2.1 m & CASLEO & \citet{Gil-Hutton2008} & 14 \\
\hline
no name & 0.9 m & McDonald & \citet{Breger1979}& 12\\
\hline
no name & 1.6 m & OMM & \citet{Angel1970}& 3\\
\hline
Hatpol & 3.8 m UKIRT & Maunakea & \citet{Hough1991}& 6\\
\hline
Vatican Polarimeter & 1.8 m & Lowell & \citet{Magalhaes1984}& 2 \\
\hline
Minipol & 1, 1.5, 2.2 m & Univ. of Arizona & \citet{Frecker1976}& 1\\
\hline
Hall & 1.06 m & Lowell&\citet{McMillan1976}& 1 \\
\hline
no name & 2.1, 4 m & Kitt Peak & \citet{Kinman1974} & 0\\
\hline
\end{tabular}
\caption{Instrument/Telescope information. (1): \url{www.not.iac.es/instruments/alfosc/polarimetry/}. Abbreviations: AIMPOL - ARIES IMaging POLarimeter; ARIES - Aryabhatta Research Institute of Observational Sciences; CTIO - Cerro Tololo InterAmerican Observatory; IAG - Instituto de Astronom\'{i}a, Geof\'{i}sica e Ci\^{e}ncias Atmosf\'{e}ricas; OPD/LNA - Pico dos Dias Observatory, Laborat\'{o}rio Nacional de Astrof\'{i}sica; HIPPO - HIgh speed Photo-POlarimeter; SAAO - South African Astronomical Observatory; AAT - Anglo-Australian Telescope, HIPPI - HIgh Precision Polarimetric Instrument; NOT - Nordic Optical Telescope; ORM - Roque de los Muchachos; WHT - William Herschel Telescope, WIYN - Winsconsin-Indiana-Yale-NOAO; FORS - FOcal Reducer and low dispersion Spectrograph; VMI - Virginia Military Institute; UCTP - University of Cape Town polarimeter; OMM - Observatoire du Mont M\'{e}gantic, CASLEO - Complejo Astron\'{o}mico El Leoncito; CAO - Crimean Astrophysical Observatory; KVA - Kungliga Vetenskapsakademien telescope; UKIRT - United Kingdom Infrared Telescope.}
\label{tab:instruments}
\end{table*}


\section{Data products}
\label{sec:products}

We provide the following data products\footnote{The products are temporarily hosted on \url{https://github.com/ginleaf/starpol\_compilation} and will be made available on \url{cdsarc.u-strasbg.fr/}.} in the form of plain ASCII files (comma-separated-value format, CSV).

\subsection{Data product 1:  {Extended catalog}}
This catalog contains the polarization measurements of all stars, stellar identification information and metadata on the original catalog from which the data were extracted. Each line contains a single measurement of fractional linear polarization and  {PA}, along with relative linear Stokes parameters and associated uncertainties.  {\textit{Gaia} EDR3 source information is included for stars that returned a match. The catalog also includes distance information for each \textit{Gaia} source from \citet{Bailer-Jones2021}. 
}

If a star was measured in several filters, each measurement is provided in a separate line.  {Thus, this} catalog contains multiple measurements of certain stars. The catalog contains 55,742 rows.
We describe the columns contained in the  {Extended catalog} below  {(see also table \ref{tab:DP1})}.

\setcounter{table}{3}
\begin{deluxetable*}{cccc}
\tablecaption{Extended catalog (abbreviated) \label{tab:DP1}}
\tablewidth{700pt}
\tabletypesize{\scriptsize}
\tablehead{
\colhead{Column number} & \colhead{Label} & \colhead{units} & \colhead{Description}
}
\startdata
1 & starID & --- & Identifier number\\
2 & Name & --- & Common  {star} name \\
3 & rawstarID & --- & Raw identifier \\
4 & RefID & --- & Reference identifier \\
5 & Instrument & --- & Instrument used \\
6 & JD & d & Julian date \\
7 & FilterID & --- & Filter identifier \\ 
8 & IntrPol & --- & Intrinsic Polarization flag \\
9 & RA & deg & Right Ascension (J2000) - approximate\\
10 & Dec & deg & Declination (J2000) - approximate\\
11 & p & --- & Fractional linear polarization \\
12 & e\_p & --- & Uncertainty in $p$ \\
13 & p\_d & --- & Debiased $p$ \\
14 & {PA} & deg & Polarization angle \\ 
15 & e\_{PA} & deg & Uncertainty in  {PA} \\
16 & q & --- & Stokes q \\
17 & e\_q & --- & Uncertainty in Stokes q \\
18 & u & --- & Stokes u \\
19 & e\_u & --- & Uncertainty in Stokes u \\
20 & EDR3\_source\_id & --- & Gaia EDR3 source id\\
21 & RA2016 & deg & Right Ascension from Gaia EDR3 (J2016)\\
22 & Dec2016 & deg & Declination from Gaia EDR3 (J2016)\\
23 & RA2000 & deg & Right Ascension from Gaia EDR3 converted to J2000\\
24 & Dec2000 & deg & Declination from Gaia EDR3 converted to J2000\\
25 & G & mag & Gaia G-band magnitude\\
26 &  r\_med\_photogeo & pc & Median estimate of the photogeometric distance\\
27 & r\_lo\_photogeo & pc & 16th percentile of the photogeometric distance\\
28 & r\_hi\_photogeo & pc & 84th percentile of the photogeometric distance\\
29 & r\_med\_geo & pc & Median estimate of the geometric distance\\
30 & r\_lo\_geo & pc & 16th percentile of the geometric distance\\
31 & r\_hi\_geo & pc & 84th percentile of the geometric distance\\
\enddata
\tablecomments{Table 4 is published in its entirety in the electronic 
edition of the  {journal}.  }
\end{deluxetable*}

\begin{deluxetable*}{cccc}
\tablecaption{Unique Source catalog (abbreviated) \label{tab:DP2}}
\tablewidth{700pt}
\tabletypesize{\scriptsize}
\tablehead{
\colhead{Column number} & \colhead{Label} & \colhead{units} & \colhead{Description}
}
\startdata
1& EDR3\_source\_id & --- & \textit{Gaia} source identifier\\
2 & starID & --- & Star identifier in the Extended catalog\\
3 & RA2000 & deg & Right Ascension (J2000)\\
4 & Dec2000 & deg & Declination (J2000)\\
5 & G & mag & G-band magnitude\\
6 & rawstarID & --- & Star identifier in the original catalog\\
7 & RefID & --- & Reference identifier\\
8 & FilterID & --- & Filter identifier\\
9 & IntrPol & ---& Intrinsic polarization flag\\
10 & p & --- & Fractional linear polarization of the maximum SNR measurement\\
11 & e\_p & --- & Uncertainty on p\\
12 &  {PA} & deg & Polarization angle\\
13 & e\_{PA} & deg & Uncertainty on the  {PA}\\
14 & r\_med\_photogeo & pc & Median estimate of the photogeometric distance\\
15 & r\_lo\_photogeo & pc & 16th percentile of the photogeometric distance\\
16 & r\_hi\_photogeo & pc & 84th percentile of the photogeometric distance\\
17 & r\_med\_geo & pc & Median estimate of the geometric distance\\
18 & r\_lo\_geo & pc & 16th percentile of the geometric distance\\
19 & r\_hi\_geo & pc & 84th percentile of the geometric distance\\
\enddata
\tablecomments{Table \ref{tab:DP2} is published in its entirety in the electronic 
edition of the  {journal}. }
\end{deluxetable*}

\subsubsection{Star identifier}
We specify a unique integer for each star provided in a given catalog, which we denote \texttt{starID}. If multiple polarization measurements of the star were provided in the original publication, each of these measurements will have the same \texttt{starID} (but will be presented in a separate line of the table). Note that if two separate publications contain measurements of the same star, these will appear in our compilation with different values of \texttt{starID}.

\subsubsection{Star name}
We retain information on the stellar identification in standard catalogs such as HD, BD, Hipparcos and \textit{Gaia} (various data releases) as a string value in the column \texttt{Name}, if available from the original publication. We use the prefix `HIP\_' for Hipparcos sources, `GaiaDR2\_' for \textit{Gaia} DR2 and `GaiaEDR3\_' for \textit{Gaia} EDR3 sources.

\subsubsection{Original catalog star identifier}
Each star is assigned a string identifier, \texttt{rawstarID}, if given in the original publication. This can be, for example, the row number of the table in which the measurement was found. If no identifier was given, but a stellar name was, we set \texttt{rawstarID} to NAN.

\subsubsection{Reference identifier}
An integer identification number, \texttt{RefID}, is used to specify the publication from which each measurement was obtained. An index is provided in Table \ref{tab:references} providing the corresponding references. 

\subsubsection{Instrument name}
A string specifies the polarimetric instrument used for obtaining a given measurement, \texttt{Instrument}. An index is provided in Table \ref{tab:instruments} with instrument names, telescope size and location, reference to a publication describing the instrument and a list of publications in our compilation that used the instrument (\texttt{RefID}s). For each instrument, a primary reference is given (usually the paper describing the instrument).

\subsubsection{Julian date}
A Julian date is given for each measurement (floating point number), \texttt{JD}, corresponding to the earliest observing date on which the observations were conducted in each of the original catalogs. For some catalogs, separate \texttt{JD}s are given for each measurement, which we include here. If no date information is given, \texttt{JD} is set to NAN.

\subsubsection{Filter identifier}
An integer identification number, \texttt{FilterID}, is assigned for the observing filter used in the original catalog. An index of all filters used and their corresponding \texttt{FilterID} and $\lambda_{\rm eff}$ is provided in Table \ref{tab:filters}.

\subsubsection{Intrinsic polarization flag}
Certain sources may show polarization that is not purely due to the ISM. We use the flag \texttt{IntrPol} to distinguish sources that are known to be intrinsically polarized. A value of 1 indicates that the original publication found signs of intrinsic polarization for a given star. Sources that are not instrinsically polarized or not known to be intrinsically polarized have \texttt{IntrPol} = 0. We caution that a value of 0 does not mean that a source is not intrinsically polarized, only that the original publication did not comment on it having polarization of intrinsic origin. 

\subsubsection{Right Ascension, Declination}
Stellar coordinates are given in the ICRS reference frame in columns named \texttt{RA, Dec}. The entries are floating point numbers in units of degree. These coordinates are in epoch J2000  {(see Section \ref{sec:homogenization} and also appendix \ref{appendix:catalogs} }for comments on how those were obtained). Values can be NAN if no stellar identification  {was} found.  {Users should take note that these coordinates in the Extended catalog are approximate, and should use the coordinates of the Unique Source catalog for most use-cases (unless a source match was not found within \textit{Gaia} EDR3).}

\subsubsection{Fractional linear polarization and uncertainty}
The fractional linear polarization and its uncertainty are provided as floating point numbers, \texttt{p, e\_p}. Values for  {these quantities} are fractional, not percentages, and lie in the range [0, 1]. The polarization fraction is a biased quantity (see Section \ref{sec:data}). Users may choose to correct the reported polarization fractions for positive bias in different ways \citep{Vaillancourt2006, Plaszczynski2014}. We provide the modified asymptotic estimator defined by \citet{Plaszczynski2014} in  a separate column.

Some catalogs \citep[e.g.][]{Berdyugin2014AA...561A..24B} only provide de-biased $p$, and have set values of $p_d$ = 0 for low SNR measurements. In these cases $p$ cannot be recovered and so we set \texttt{p} to NAN. If no uncertainty is provided for $p$, we set \texttt{e\_p} to NAN. 
In some catalogs with multi-band measurements, it is common to not have measurements for each source in all observing bands. In these cases polarization measurements have been reported as NAN values in the original catalogs. We discard such entries from entering our compilation. 

\subsubsection{De-biased fractional linear polarization}

Because $p$ defined in equation \ref{eqn:evpa} is a positive-definite quantity, it provides a biased estimate of the true fractional linear polarization of a source for small signal-to-noise ratios \citep{Vaillancourt2006}. We provide the de-biased fractional linear polarization $p_d$ \citep{Plaszczynski2014} as defined in equation \ref{eq:pd}, for all measurements where $e_p$ is also available. As for $p$ and $e_p$, $p_d$ is expressed as a fraction, not as a percentage.

\subsubsection{Polarization position angle and uncertainty}
The {PA} and its uncertainty are given in degrees (\texttt{{PA}, e\_PA}). They are defined in the  {equatorial} celestial frame, following the IAU standard convention \citep{Contopoulos1974}: increasing East from North. The  {PA} is within the range  {[0, 180$^\circ$]}. When computing the  {PA} from Stokes parameters, we use eq. \ref{eqn:evpa} (note the use of the 2-argument artangent, which returns angles in the full range of interest). 
We compute \texttt{e\_pa} by eq. \ref{eqn:sigmatheta}. 
Note that our estimate of the  {PA} uncertainty is tied to our choice of de-biased $p$ estimator. If no uncertainty of $p$ is given, we set \texttt{e\_pa} to NAN.
There may be systematic differences in the  {PA} among catalogs due to the  {coordinate reference frame (epoch of observation, see Section \ref{sec:homogenization})}.

\subsubsection{Relative linear Stokes parameters}
The relative linear Stokes parameters \texttt{q, u} and their uncertainties \texttt{e\_q, e\_u} are defined in the  {equatorial} celestial reference frame according to the IAU convention (see Section \ref{sec:homogenization_p} on {PA}). They are fractional values, in the range [-1,1]. 

\subsubsection{Gaia EDR3 source information}
 {The column \texttt{EDR3\_source\_id} holds the \textit{Gaia} EDR3 source identification number of the best match. 
The columns \texttt{RA2016, Dec2016} specify the celestial coordinates (J2016) of the \textit{Gaia} match, while \texttt{RA2000, Dec2000} contain coordinates in J2000. The column \texttt{Gmag} contains the G-band apparent magnitude. Columns \texttt{r\_med\_photogeo}, \texttt{r\_lo\_photogeo}, \texttt{r\_hi\_photogeo} contain the distance estimate as well as the lower and upper 1-$\sigma$ bounds on the distance from the photo-geometric method described in \citet{Bailer-Jones2021} (see also Section \ref{sec:gaia}). These estimates are to be preferred for most sources. Columns \texttt{r\_med\_geo}, \texttt{r\_lo\_geo}, \texttt{r\_hi\_geo} contain the same distance estimate information, but for the geometric method used by those authors.}


\begin{figure*}
    \centering
    \includegraphics[scale = 0.9]{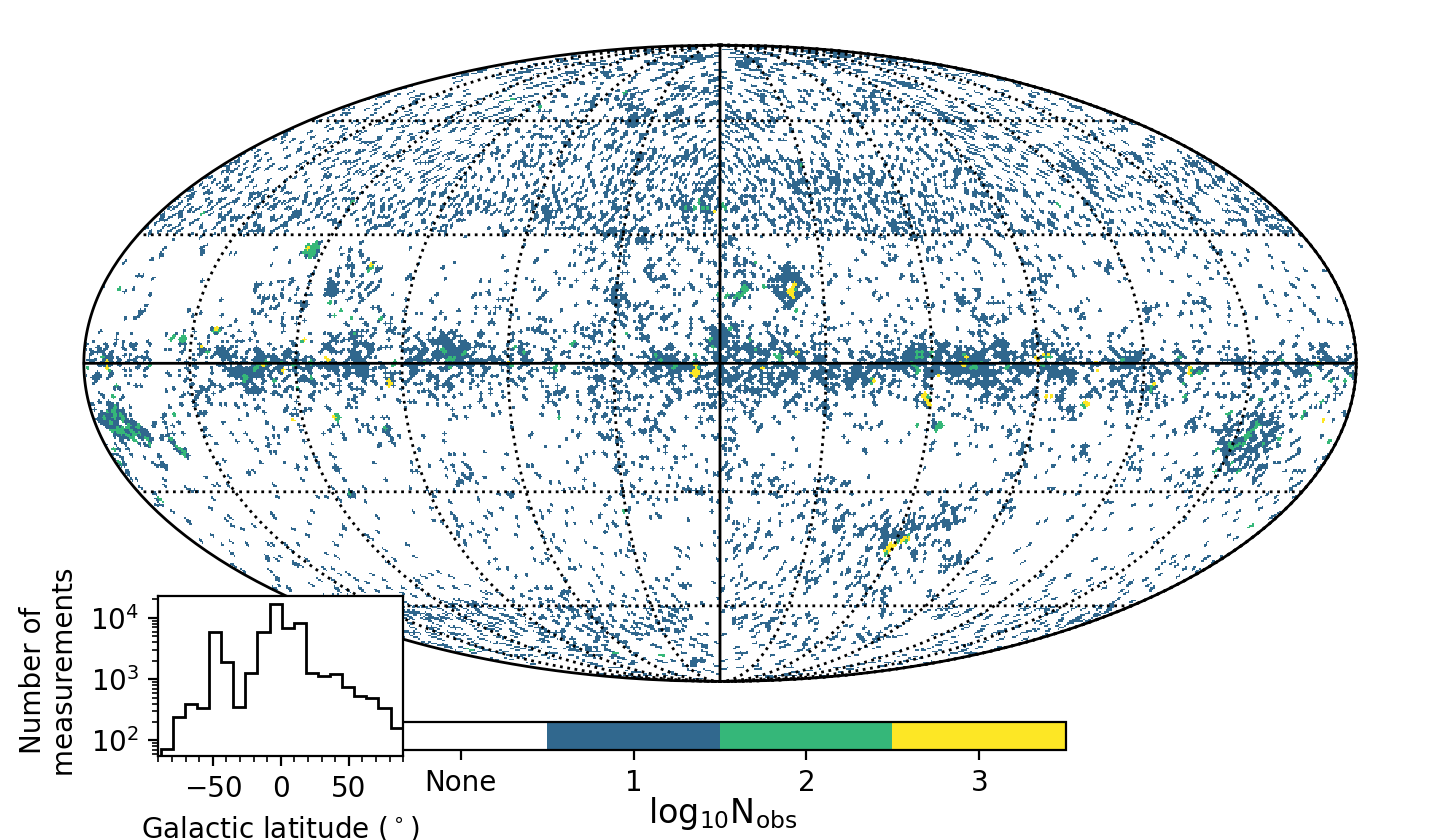}
    \caption{Map of the number of polarization measurements in the catalog per Nside = 64 pixel ($\sim 1^\circ$) in Mollweide projection. The map is centered on (l,b) = (0$^\circ$,0$^\circ$). White pixels contain no measurements. The inset shows the distribution of galactic latitudes of the stellar polarization measurements on a semi-logarithmic axis.
    }
    \label{fig:Nobs}
\end{figure*}

\subsection{Data product 2: Unique Source catalog}

 {Data product 1 contains} multiple entries for certain stars. We produce a  {second} catalog that contains only unique \textit{Gaia} sources. 
For each \textit{Gaia} EDR3 source  {in the Extended catalog}, we search for \texttt{starID}s that correspond to it. 

If there are multiple polarization measurements corresponding to a given \textit{Gaia} source, we keep the measurement with highest SNR in $p$ ($p/e\_p$). Thus, if a source has measurements in multiple filters, we retain only one measurment (from whichever filter provided the highest SNR datapoint). If all matching \texttt{e\_p} are NAN, we keep the measurement with highest $p$. 

 {The Unique Source catalog contains 42,482 unique stars. 
It includes unique values of \texttt{starID}. A certain source has the same \texttt{starID} in both the Extended catalog and in the Unique Source catalog.
}
 {Table \ref{tab:DP2} presents column information for the Unique Source catalog.}
 {All column names are defined as for the first data product}. 

The Unique Source catalog demonstrates one way in which a catalog with no duplicate source entries can be constructed. In this specific case, polarization measurements have been selected to have high SNR, but different wavelengths of observation (filters).  {We caution users to take into account the wavelength dependence of polarization \citep{Serkowski1975} and how it may impact their analysis when data from multiple filters are combined.}  Depending on the usage-case, other unique source catalogs may be constructed from the information contained in the  two data products  {described above}.

\subsection{Filter index and Reference index}
 Table \ref{tab:references} specifies the publication that corresponds to each value of the reference identifier (\texttt{RefID}). Table \ref{tab:filters} second contains information on all filters used in the original catalogs. Table \ref{tab:filters}  is provided in machine-readable format in the journal. The columns are: Filter name, unique filter identifier used in the  {Extended} catalog, and central wavelength in microns.

\section{Overview of the data}
\label{sec:results}

We visualize the area coverage of the measurements in the Extended catalog by constructing a HEALPix \citep{Gorski2005ApJ...622..759G} map at $N_{\rm{side}} = 64$ (pixel size of 55\arcmin). Figure \ref{fig:Nobs} shows a map of the logarithm of stellar polarization measurements per 55\arcmin pixel in Galactic coordinates. The majority of measurements are in the Galactic plane. There are 9423 pixels with at least one stellar polarization measurement, corresponding to 7908 square degrees, or 14\% of the sky. The number of measurements exceeds 10 per 55\arcmin pixel in only 290 pixels. Considering only pixels with at least 1 measurement, the mean number of measurements per pixel is 2.


Figure \ref{fig:phist}  {(left)} shows the distribution of $p_d$ in percentage for all measurements in the Extended catalog. The 10- and 90-percentiles of the distribution of $p_d$ are 0.2\% and 4.6\%, respectively, while the median is 1.6\%.
There exist sources with very high polarization, unlikely to be entirely of interstellar origin. Of the 12 highest polarization sources (those with $p>20\%$) only two are flagged as intrinsically polarized based on the original publications. The remaining ten are all in the sample of \citet{Santos2014ApJ...783....1S} towards the star-forming region Sh 2-29 and could be intrinsically polarized, but no such mention was found in the publication. The authors claim that no significant deviation of the  {PA} is observed over the entire sample, which indicates that intrinsically polarized sources should be few, if any.

 {Figure \ref{fig:phist} (right) shows the distribution of SNR of $p$ ($p/e\_p$) for measurements in the Extended catalog. The majority of measurements are significant detections, with 80\% of data having SNR $> 3$. Since the polarization data arise from a very inhomogeneous set of observing strategies, instruments and selection criteria, we caution that any selection based on e.g. an SNR threshold may introduce further biases in the sample. }

Figure \ref{fig:G-dist} (left) shows the apparent brightness of all \textit{Gaia} EDR3 sources in the  {Unique Source catalog}. The distribution shows two main peaks (at $\sim 8.5$ and $\sim 17.5$ mag). These two peaks correspond to telescope specifications and survey strategies that impose observational constraints. The brighter source peak is comprised of data from \citet{Heiles2000AJ....119..923H}, as well as from telescopes with apertures smaller than 0.9 m or from instruments that primarily target bright stars \citep[e.g.][]{Bailey2015}. Not surprisingly, these inhomogeneous brightness limits also translate to a double-peaked distance distribution, seen in Figure \ref{fig:G-dist} (right). For the distribution of distances, we have selected only sources with relatively high SNR in distance. To make this selection, we define a symmetric distance error of: \begin{equation}
    \begin{split}
        \rm r_{\rm err} = 
         max( & \texttt{r\_med\_photogeo} - \texttt{r\_lo\_photogeo}, \\
         & \texttt{r\_hi\_photogeo} - \texttt{r\_med\_photogeo})
    \end{split}
\end{equation} and we choose sources with:
\begin{equation}
    \frac{ \texttt{r\_med\_photogeo} }{ \rm r_{\rm err}} > 3.
\end{equation}
We also exclude sources that are part of the \citet{Lobo2015ApJ...806...94L} catalog, as the Galactic distance prior employed by \citet{Bailer-Jones2021} does not apply for stars in the Magellanic system.
The bulk of stars in the  {Unique Source catalog} ($\sim 70$\%) that satisfy the aforementioned constraints are within 2 kpc from the Sun. 

\begin{figure*}[ht]
    \centering
    \includegraphics[scale = 1]{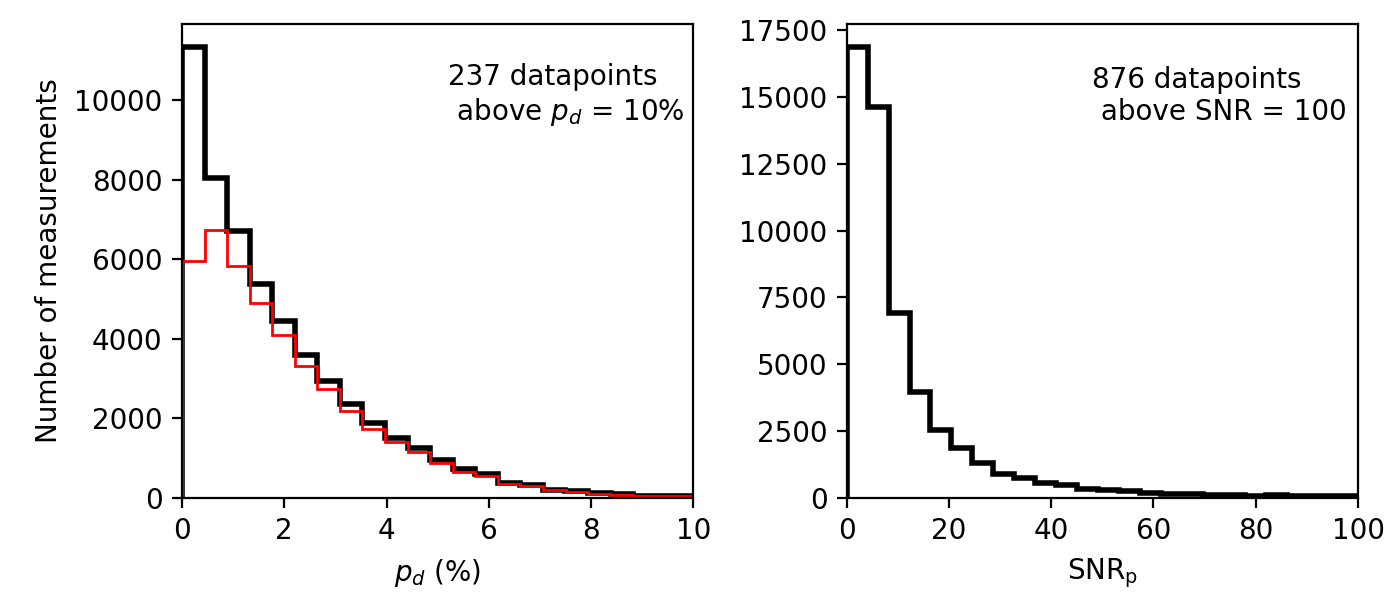}
    \caption{Polarization properties of measurements in the Extended catalog. Left: Distribution of de-biased polarization percentage for all measurements (black) and for significant detections (SNR$_p \geq 3$) (red). Right: distribution of SNR of $p$ for all measurements with non-NAN $e_p$.}
    \label{fig:phist}
\end{figure*}

\begin{figure*}[ht]
    \centering
    \includegraphics[scale = 1]{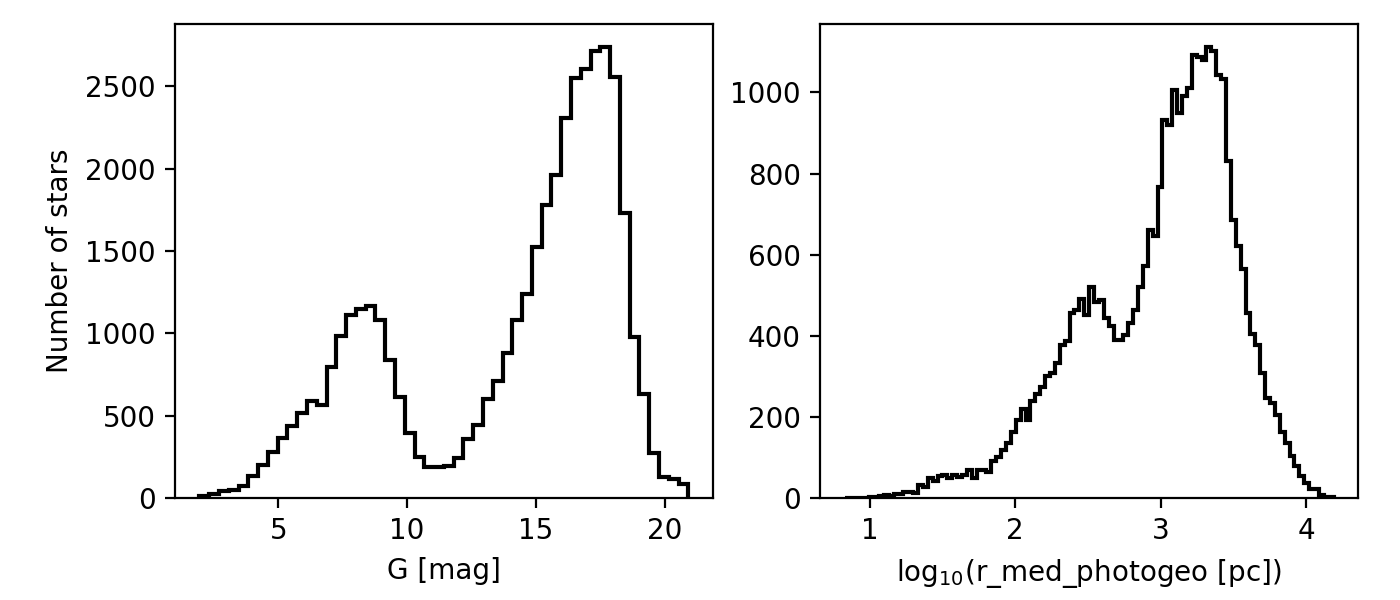}
    \caption{Properties of \textit{Gaia} EDR3 sources in the  {Unique Source catalog}. Left: Distribution of G-band magnitude. Right: Distribution of the distance estimate \texttt{r\_med\_photogeo} for sources with SNR in distance $> 3$ and not in the \citet{Lobo2015ApJ...806...94L} catalog (see text).}
    \label{fig:G-dist}
\end{figure*}

\begin{figure*}[ht]
    \centering
    \includegraphics[scale = 1]{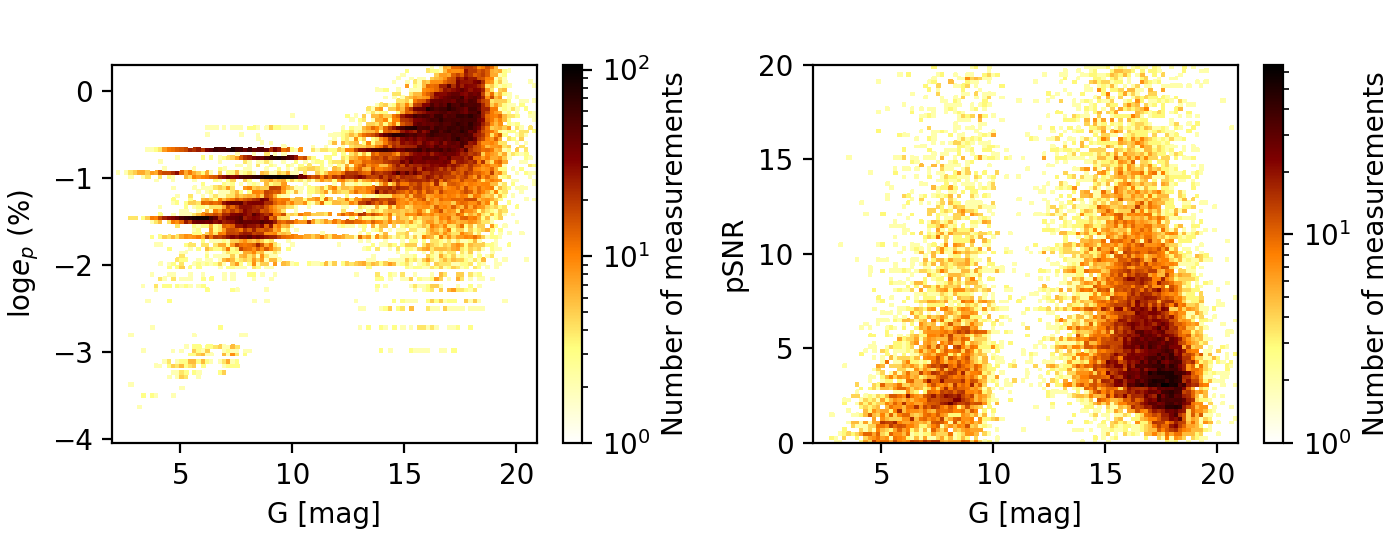}
    \caption{Polarization properties of measurements in the  {Unique Source catalog}. Left: 2-d distribution of the logarithm of $e_p (\%) $ versus G-band magnitude. Right: 2-d distribution of SNR of $p$ versus G-band magnitude. In both panels the colorscale is logarithmic.}
    \label{fig:epGsnr}
\end{figure*}

\begin{figure}[ht]
    \centering
    \includegraphics[scale = 0.9]{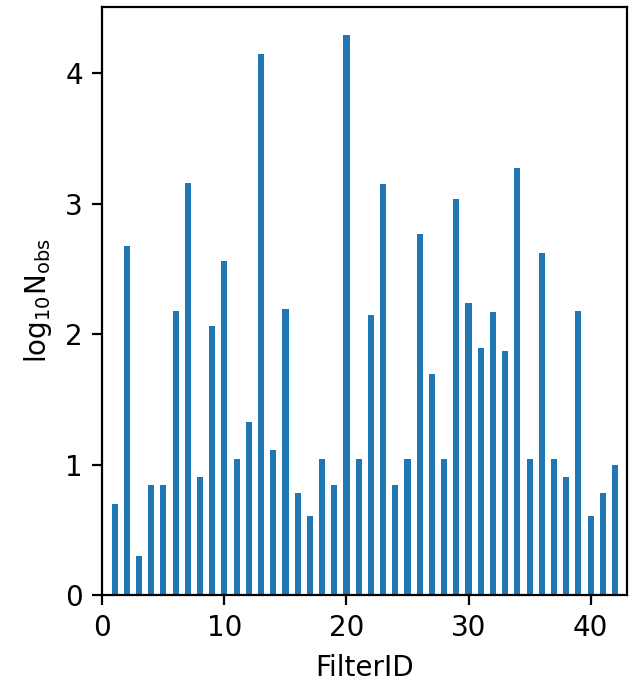}
    \caption{Distribution of measurements in the various filters within the Extended catalog. The vertical axis shows the logarithm of the number of datapoints in each filter (specified by the FilterID, see table \ref{tab:filters}).}
    \label{fig:filt}
\end{figure}

 As noted previously, the compilation comprises of data from various surveys with varying observing strategies, sensitivities etc. The inhomogeneity of polarimetric accuracy of the data is clearly seen in Figure \ref{fig:epGsnr} (left), where we show the 2-d distribution of the logarithm of the uncertainty in $p$ versus G-band magnitude. The envelope of log$e_p$ rises with G, as expected for magnitude-limited surveys (fainter sources are less-well measured). The distribution is not smooth, with a gap around G of 10 mag. The distribution also exhibits discontinuities, arising from the finite precision (decimal places) with which the errors were published in some catalogs. The inhomogeneous accuracy of the data is also seen in the 2-d distribution of SNR in $p$ versus G (Figure \ref{fig:epGsnr}, right). The two main sub-samples of stars (G$<10$ and G$>10$) have differently distributed SNRs.

 {
Figure \ref{fig:filt} shows the number of measurements ($N_{obs}$) per filter in the Extended catalog. The majority of data are in the R and V bands (FilterIDs 20 and 13).
}

\subsection{Duplicate entries}
\label{sec:duplicates}

 {The Extended catalog may contain multiple entries for any given star.} This is the case, for example, if multiple measurements of a star at different wavelengths or at different times are presented in a publication. In this case, all these measurements in our catalog will have the same \texttt{starID}. Multiple (i.e. duplicate) entries can also exist if the same star has been measured in multiple catalogs (from different publications). 
The measurements may not be independent, as can happen either when an earlier measurement is reported as-is in a later publication $-$ for example \citet{Berdyugin2014AA...561A..24B} includes some of the \citet{Berdyugin2002AA...384.1050B} data. 
The more common situation is when new observations of a star with already published polarization are conducted (in the same band). In these cases the multiple measurements of the star could be combined to obtain an updated estimate. However this is not advised, for a number of reasons. First, the two measurements could be inconsistent, as a result of (a) intrinsic variability of the polarization of the star or (b) unaccounted-for systematic uncertainties in any of the observations.  {If different filters were used, then the wavelength-dependence of polarization \citep{Serkowski1975} should be taken into account when attempting to combine the data.} We note that multiple measurements could be erroneously assigned to the same \textit{Gaia} source, as a result of mis-identification in the cross-matching (e.g. in crowded fields).

We can use such duplicate measurements to comment on the consistency of the data across different catalogs. Figure \ref{fig:duplicates} shows a comparison between measurements that are common in the \citet{Heiles2000AJ....119..923H} catalog and other catalogs. The top panel presents differences of the polarization fraction $\Delta p  = p_{Heiles} - p_{catalog_2}$ relative to their uncertainty $ e_{\Delta p}$. In this notation, $catalog_2$ represents the comparison catalog other than \citet{Heiles2000AJ....119..923H}. Individual measurements are shown in blue. The median $\Delta p /e_{\Delta p}$ is shown for each catalog separately in black. We mark the 15- and 84- percentiles of the distribution of relative differences for each catalog as black error-bars.
The median relative polarization fractions are within 1.1$\sigma$ of the \citet{Heiles2000AJ....119..923H} data points, however there are noticeable outliers. We caution that the exact $\lambda_{\rm eff}$ was not determined for the \citet{Heiles2000AJ....119..923H} (see relevant discussion in Appendix \ref{appendix:catalogs}) and thus the comparison presented here does not necessarily correspond to measurements performed with the same filter. Some variation is therefore expected. 

The bottom panel of Fig. \ref{fig:duplicates} shows the relative  {PA} difference, $\Delta {PA} / e_{\Delta {PA}}$, where $\Delta {PA} = \theta_{Heiles} - \theta_{catalog_2}$. Again, the median relative angle difference for most catalogs is consistent within 1$\sigma$ of the \citet{Heiles2000AJ....119..923H} data. 
However, there are some cases where the differences are large. We note the significant discrepancies in the comparison to the \citet{Santos2011ApJ...728..104S} data (catalog \texttt{RefID} = 29). There are 60 stars in common between the \citet{Santos2011ApJ...728..104S} and \citet{Heiles2000AJ....119..923H} catalogs. All are bright stars with G $<$ 9 mag with known Hipparcos identifiers, thus mis-identifications are less likely to have occurred. We have checked that the 10 most discrepant measurements are not mis-identifications or transcription errors by verifying with the original \citet{Mathewson1970} catalog. 
We note that while all highly discrepant measurements in  {PA} are highly significant measurements of the polarization, the quoted uncertainties are quite low $e\_p < 0.06\%$. While some of the observed discrepancies between the measurements of the two catalogs could be due to intrinsic variability of stars, it is unlikely that all 60 stars show variability only in  {PA} and not in p. This may indicate problems in the absolute calibration of the angles in the \citet{Santos2011ApJ...728..104S} study.

\begin{figure}
    \centering
    \includegraphics[scale = 0.95]{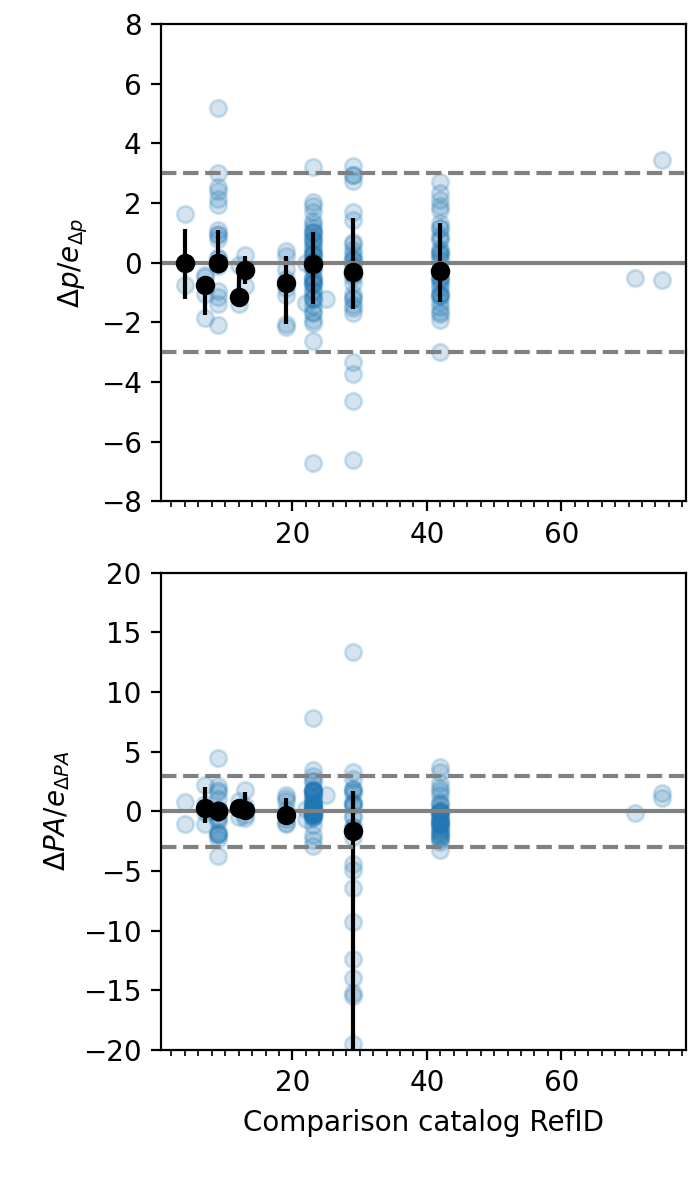}
    \caption{Consistency of star polarization measurements for stars in common between \citet{Heiles2000AJ....119..923H} and other catalogs. Top: Relative difference between the polarization fraction measured in two separate catalogs, as a function of the comparison catalog \texttt{RefID}. Bottom: same but for relative difference of the polarization angle. Blue circles mark individual stars, while black points mark the median value of the distribution of relative differences for stars in a specific comparison catalog. Errorbars denote the 15- and 84- percentiles of the distribution of differences per comparison catalog. Note the large discrepancies between  {PA} measured in catalog 29 \citep{Santos2011ApJ...728..104S} compared to \citet{Heiles2000AJ....119..923H}.}
    \label{fig:duplicates}
\end{figure}

\section{Recommendations}
\label{sec:future}

\subsection{For users}

Users must be aware of the presence of unflagged intrinsically polarized sources. It is highly likely that the catalog contains sources that are intrinsically polarized, but have not been identified as such by the authors of each original catalog. We remind readers of the existence of duplicate measurements per source. We have presented one way of creating a catalog with a single measurement per source (Data product  {2}), but there are many different options one could select instead.

We note that combining multiple measurements of the same star (e.g. via weighted average of the Stokes parameters) should be done with caution. Polarization measurements can be discrepant between different studies (see Section \ref{sec:duplicates}). 
 {We remind readers of the wavelength dependence of interstellar polarization \citep{Serkowski1975}, which should be kept in mind when combining data from different filters.}

We strongly encourage users of the compilation to cite the original publications of the data they use. The references can be imported into bibtex via the dedicated ADS library,  found at \url{https://ui.adsabs.harvard.edu/public-libraries/G7aXNeaxTjqL1zyMVofxKw}.

\subsection{For future catalogs}


As more data become available from a variety of groups around the world, maintaining a common database and bibliographic record will facilitate discoverability. For this reason, we propose the use of an ADS bibliographic library that can be updated as new data relevant to stellar polarization and ISM studies are published. 

We encourage the adoption of a common tabular format for the publication of stellar polarization data. This would greatly facilitate the collection of future datasets and dramatically reduce the time needed to create a homogeneous catalog such as the one presented here. The tabular formats presented in Table \ref{tab:DP1} and Table \ref{tab:DP2} can serve as a guide. We envision the adoption of such a standard would greatly benefit the community, and allow for wider dissemination of scientific results, in the spirit of that proposed by \citet{VanEck2023} for rotation measure data.

\section{Summary}

We have compiled a new optical polarization catalog by combining data from 81 publications. We have presented a homogeneous stellar polarization catalog with 55,742 measurements (Data product 1). We have cross-matched this catalog with \textit{Gaia} EDR3  {and have presented} a catalog of combined polarization and stellar distance information for a unique set of sources in \textit{Gaia} EDR3 (Data product  {2}). 

To facilitate the discoverability and collection of stellar polarization data in the future, we propose the use of a common tabular format, as presented in Data products 1 and  {2}. We hope that our compilation will help in standardizing the publication of similar datasets, and that it will be of use to the ISM community at large.





\acknowledgments

We thank C. Eswaraiah and A. Chakraborty for sharing machine-readable versions of their data in Bijas et al 2022 and Chakraborty et al 2016, respectively. We thank A. Goodman and G. Topasna for providing clarifications on their datasets. D.B. and V.P. acknowledge support from the European Research Council (ERC) under the European Union Horizon 2020 research and innovation program under the grant agreement No 771282. This research has made use of the WEBDA database, operated at the Department of Theoretical Physics and Astrophysics of the Masaryk University. This work has made use of data from the European Space Agency (ESA) mission
{\it Gaia} (\url{https://www.cosmos.esa.int/gaia}), processed by the {\it Gaia}
Data Processing and Analysis Consortium (DPAC,
\url{https://www.cosmos.esa.int/web/gaia/dpac/consortium}). Funding for the DPAC
has been provided by national institutions, in particular the institutions
participating in the {\it Gaia} Multilateral Agreement. This work made use of astropy:\footnote{http://www.astropy.org} a community-developed core Python package and an ecosystem of tools and resources for astronomy \citep{astropy:2013, astropy:2018, astropy:2022}. This research has made use of "Aladin sky atlas" developed at CDS, Strasbourg Observatory, France \citep{Aladin}. This research has made use of NASA’s Astrophysics Data System.



\software{astropy \citep{astropy:2013}, astroquery \citep{Ginsburg2019AJ....157...98G}         }



\clearpage

\appendix

\section{Notes on treatment of individual catalogs} \label{appendix:catalogs}

\subsection{RefID 4: Heiles 2000}

The most well-cited compilation of catalogs is that of Heiles \citep{Heiles2000AJ....119..923H}, who gathered data for 9286 stars from 11 catalogs. The catalog contains linear polarization measurements, V-band photometry, spectral types and distances from Hipparcos \citep{Perryman1997AA...323L..49P} for a large number of entries. 

The observations in \citet{Heiles2000AJ....119..923H} were collected from various investigations each using a different instrumental setup, resulting in a variety of effective wavelengths. For stars with multiple observations (from different catalogs, at different wavelengths) the reported measurements of polarization in \citet{Heiles2000AJ....119..923H} correspond to weighted averages over all these measurements. For these reasons we have not been able to specify a unique $\lambda_{\rm eff}$ for the data in \citet{Heiles2000AJ....119..923H}, and have therefore set \texttt{FilterID} = 0. 

We have identified the following errors in the values/stellar identifications in this catalog, which we have corrected. On a few occasions the HD number was incorrect:
HD 72317 was the wrong HD number for the star, which was really
CD -39.10433 (the given CD number was also incorrect). Our determination agrees with the original publication from which the data were taken. Some stars had incorrect HD numbers and we were unable to find their correct matches. They were: HD 31243, HD 14615, HD 67018, HD 14520, HD 58955, HD 328864, HD 39477, HD 67018. The polarization fraction  and  {PA} are erroneously given for two stars HD 11471 and BD+60 387. The error was originally in \citet{Mathewson1970} and was propagated into \citet{Heiles2000AJ....119..923H} (we thank B-G Andersson for notifying us of this error). 
HD 11471 \citep[rawstarID = 8827 in][]{Heiles2000AJ....119..923H} has a quoted $p = 4.24 \pm 0.1$\%  and  {PA}=$102.8^\circ \pm 0.7 ^\circ$, but the original source \citep{Schmidt1968ZA.....68..380S} has $ p=1.34 \pm 0.17$\% and  {PA}=$89^\circ \pm 3^\circ$. We have used equation $p(\%) = 46.05 \cdot p_{mag}$, to transform polarization in magnitudes to polarization expressed in percentage \citep[see ][for a derivation]{Hall1958PUSNO..17..275H}. 
The values given in \citet{Mathewson1970} for BD+60 387 \citep[\texttt{rawstarID} = 8741 in][]{Heiles2000AJ....119..923H} are $p$ = 1.64$\pm$0.1\%,  {PA} = $100.7^\circ \pm 1.7^\circ$, but calculating from \citet{Schmidt1968ZA.....68..380S} yields $p$ = 1.178$\pm$0.1\% and  {PA} = $ 86.4^\circ \pm 3.0^\circ $. 

The canonical value for NAN for the polarization fraction uncertainty is -999.9 in \citet{Heiles2000AJ....119..923H}. However, we found that some measurements were provided with $e_p = 0$. We set $e_p$ to NAN for those cases as well.

\subsubsection{Preparation for cross-matching Heiles (2000) with Gaia DR2:}
For the \citet{Heiles2000AJ....119..923H} catalog, we performed a separate cross-match with Gaia, due to the fact that for many stars the coordinates were uncertain. Any stars with no good match with Simbad were excluded from the cross-match with Gaia. 

Prior to cross-matching, there were several steps needed to obtaining accurate coordinates for the stars in this catalog:
\begin{itemize}
    \item[I.] Stars with identifiers in the Henry-Draper catalog \citep{Cannon1918}: query Simbad by star identifier.
    \item[II.] Stars with IDcat = 5 (if no HD number): cone search within 1.5 arcminutes with Simbad.
    \item[III.] Stars with no HD number and IDcat$\neq 5$:  {special treatment including visual inspection.}
    \item [IV.] Stars with IDcat = -999.9: not included in the catalog (but see text for details on how some are recovered).
\end{itemize}
We explain our reasoning and treatment of special cases below. 

For all stars with an HD number, we queried Simbad using the method \texttt{Simbad.query\_objects} from the \texttt{astroquery} python package. For stars with identifications in the BD, CD, or CPD catalogs, the query often did not return results. One of the problems came from inconsistent name formats between the publication and the online service. For example, the star named BD 76 85400 in the catalog could be BD 76 854 or 76 8540, and required manual checking to ascertain which was its correct name. Of the stars with HD number, sometimes the Simbad query failed to find a match. On some occasions (mentioned above), we believe the HD numbers are wrong.

Stars with IDcat = 5 needed special treatment as their coordinates were off by up to a few arcminutes. There are 88 such stars. Of these, 21 had a HD identified which we used for querying Simbad. For the remaining 67, we queried Simbad for bright stars within 1.5 arcminutes of the coordinates given in the catalog. If there were more than 1 stars returned, we identified manually which was the correct match (combining angular distance, and visual magnitude information if present in the catalog). 

For stars with no HD number and IDcat $\neq$5, we used the coordinates as given in the Heiles catalog for cross-match with Gaia. This sometimes resulted in no good match. In some cases we were able to visually locate the correct star by looking for bright stars in the vicinity on DSS images through the \texttt{Aladin sky atlas} software \citep{Aladin}, and taking into account the visual magnitude in the catalog, if it existed, and updated its coordinates manually.

There are 114 stars for which the automated Simbad query did not return any match or returned multiple matches but we were able to recover accurate coordinates manually or determine that there was an error and a correct match could not be found even manually. 

There are 1180 stars in the Heiles catalog that are flagged as having uncertain stellar positions (IDcat = -999). We do not include the measurements as given in Heiles for any star with IDcat = -999. However, of these flagged stars, the majority (782) have been flagged as uncertain because the original dataset (all come from a private communication with Goodman et al. (1997)) did not contain stellar identifications and therefore checks could not be made on the positions. Due to the substantial number of stars in this single compilation, we attempted to recover accurate coordinates for this sample. Examination of the positions of these stars reveals that the majority are part of several dark cloud complexes. These appear to be the same complexes studied in \citet{Goodman1990ApJ...359..363G}. We therefore include the data published in \citet{Goodman1990ApJ...359..363G}, as well as data from other catalogs that are used in their paper \citep{Vrba1976AJ.....81..958V,Heyer1987ApJ...321..855H,Moneti1984ApJ...282..508M}.
By using these original datasets, we are able to recover uncertainties for the $p$ and PA, which were not included in the \citet{Heiles2000AJ....119..923H} entries. See section \ref{sec:older_catalogs} for details.

\subsubsection{Cross-matching Heiles (2000) with Gaia DR2/Hipparcos:}

Having coordinates for the majority of stars in the \citet{Heiles2000AJ....119..923H} catalog, we then proceeded to perform a cross-match with \textit{Gaia} DR2. 
For each star, we select a query radius of 10\arcsec. We query \textit{Gaia} DR2 within that radius for all stars with G $<$ 13 mag using the \texttt{gaia} class from the \texttt{astroquery} python package. If there is more than 1 match returned, we choose the brightest star (or the star with G-band magnitude closest to the given V-band magnitude given in the catalog). If no match is returned with Gaia DR2, we revert to checking if it has a match with Hipparcos. For this we query the \texttt{public.hipparcos\_newreduction} table from within the \textit{Gaia} archive. If there is no Hipparcos star in the queried area, we repeat the search for a star with a wider area by visual inspection. If there are multiple Hipparcos matches, we again solve the degeneracy by visual inspection.
Any stars that showed very discrepant G-band magnitudes compared to their original V-band values were visually inspected and corrected or set as having no good match. We did not transform the J2015 coordinates from \textit{Gaia} to J2000, as we expect positional changes less than our search radius of 5 arcseconds due to stellar proper motions. Indeed, we have calculated the angular separation between the J2015 positions and the positions propagated to J2000 for all sources with measured proper motion in Gaia DR2 and confirmed it is less than 5 arcseconds for all but a couple targets.

\begin{figure}[ht!]
    \centering
    \includegraphics{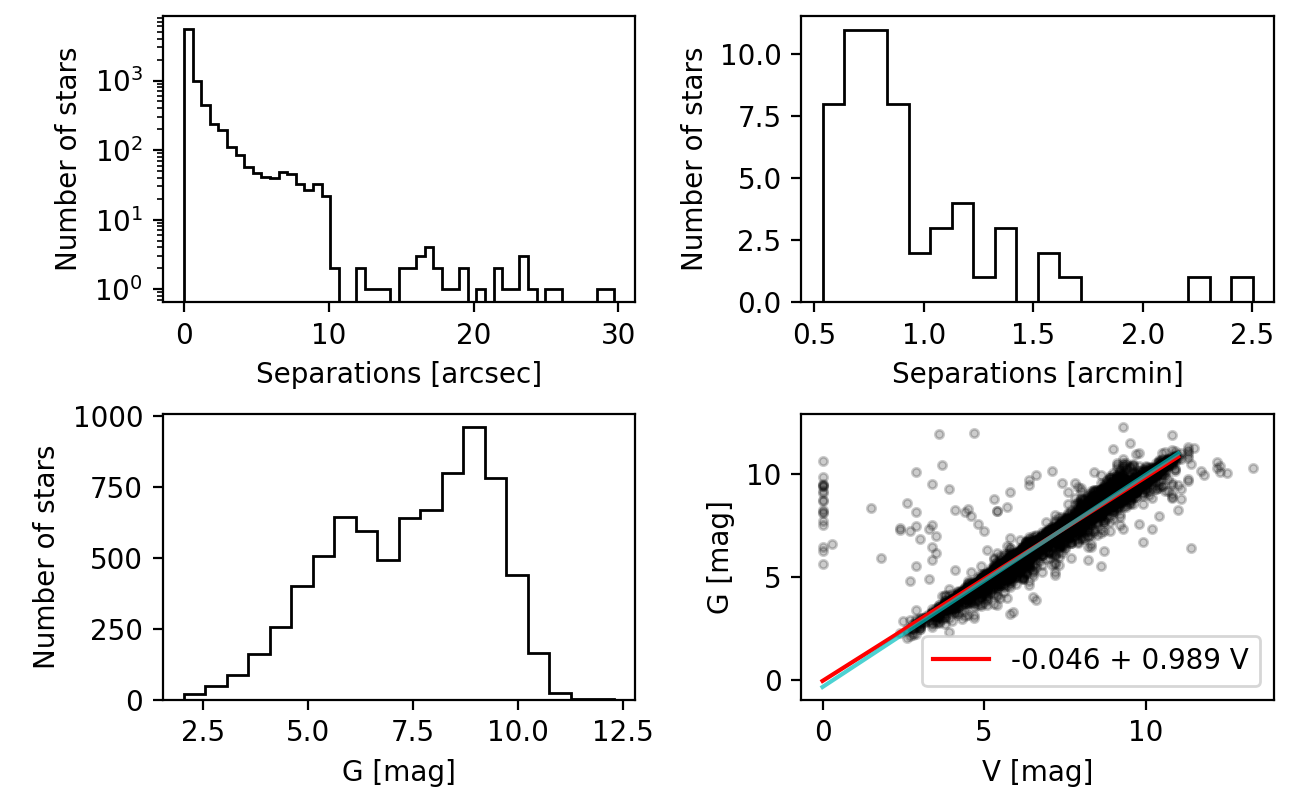}
    \caption{Properties of the sources with successful cross-matches between \citet{Heiles2000AJ....119..923H} and \textit{Gaia} DR2. Top left: Angular separations between original coordinates given in catalog and the final \textit{Gaia} matches in arcseconds. Does not include 56 sources with separations larger than 30 arcseconds. Top right: Same as top left, but showing only the 56 sources with angular separations larger than 0.5 arcminutes, note the change of horizontal axis units to arcminutes. Bottom left: Distribution of G-band magnitudes of the successful matches from \textit{Gaia} DR2. Bottom right: Comparison between \textit{Gaia} G-band magnitudes and V-band magnitudes originally reported in \citet{Heiles2000AJ....119..923H} for all sources with both G- and V- measurements. We overlay a fit to the data (red line) excluding sources with V = 0 (these are erroneous). The slope and intercept are showin in the legend. We also overlay the fit from the cross-match of \citet{Heiles2000AJ....119..923H} with Gaia DR2 from \citet{Meng2021ChAA..45..162M} (cyan line), which shows a similar slope but very different intercept (see text). }
    \label{fig:heiles}
\end{figure}

In the final cross-matched catalog with \textit{Gaia}, there are 56 stars that showed angular separations larger than 30 arcseconds between the original Heiles 2000 coorinates and the ones we assigned (of these most had separations of 1 arcminute). There are 96 that showed separations larger than 10 arcseconds. There are 2 stars that showed the largest angular separations: HD 312052 had 2.2 arcminutes separation between the original coordinates and those returned from Simbad. Star HD 144558 had 2.5 arcminutes separation from the original coordinates. We have visually confirmed that these are correct matches. 
325 stars have Hipparcos matches and no Gaia match. 

We note that at the time of performing the analysis of the \citet{Heiles2000AJ....119..923H} catalog, Gaia EDR3 had not been public. We have chosen to use the cross-match results from Gaia DR2 and simply find their corresponding \texttt{source\_ids} in Gaia EDR3 by querying the \texttt{gaiaedr3.dr2\_neighbourhood} table of the Gaia archive. This was done instead of repeating the cross-match of Heiles 2000 directly with Gaia EDR3 to avoid the lengthy process of manual checks and matches needed for many of the stars. Due to the bright nature of the sources, we believe that repeating the process with EDR3 would have yielded a similar amount of matches. The brightest stars that were missed in DR2 are included as Hipparcos matches.

Figure \ref{fig:heiles} shows the comparison between \textit{Gaia} DR2 G-band magnitudes obtained after our cross-match and checks and V-band magnitudes provided in the original catalog. The majority of sources (7938 in total) have angular separations smaller than 10 arcseconds. There remain 97 sources with larger separations, up to 2.5 arcminutes. As explained previously, we are confident that these also represent correct matches and not mis-identifications, as a result of the process and checks described above. The recovered matches are all bright (G$< 12.5$ mag), which attests to the quality of the match. The G-band magnitudes from \textit{Gaia} agree on average with the originally reported V-band magnitudes. We fit a line to the data of all sources that have both a G-band and V-band measurement (excluding Hipparcos matches and sources with V = 0 mag - erroneous measurements). We find a slope of 0.989$\pm 0.003$ and intercept of -0.046. We compare our results with the cross-match of the \citet{Heiles2000AJ....119..923H} with Gaia DR2 presented by \citet{Meng2021ChAA..45..162M}. In their analysis, a  cross-match radius of 30 arcsecond or 120 arcseconds was used, and the match was determined taking into account photometric information. We find fewer outliers than in their catalog, which may explain the much smaller intercept of -0.046 that we find compared to their value of -0.34. Notably, we find no matches with $G > 12.5$ mag, whereas their sample contains sources with $G$ up to 20 mag - the latter are clearly mis-identifications, as commented by the authors. The slope of 0.989 is within the 2-$\sigma$ uncertainty consistent with their slope of -1.03.

\subsection{RefIDs 0, 1, 2, 3: Vrba 1976, Monetti 1974, Heyer 1987, Goodman 1990}\label{sec:older_catalogs}

Data from \citet{Vrba1976AJ.....81..958V, Moneti1984ApJ...282..508M,Heyer1987ApJ...321..855H,Goodman1990ApJ...359..363G} were transcribed manually to the common tabular format used for our compilation.  

We note there is likely a typographical error in \citet{Goodman1990ApJ...359..363G} table 1: the RA for source number 4 is given as `5h 15m 38s' whereas all other stars in the table have RAs '4h XX XX' and the entire sample is said to trace the L1506 cloud. We find no bright star in the originally quoted coordinates, and the location is several degrees off of L1506, therefore we conclude that the hour entry has been given erroneously. By instead correcting the RA to `4h 15m 38s' the star is located within the area of L1506 and in the vicinity of a bright star. We have made this correction in our catalog.

The coordinates of \citet{Goodman1990ApJ...359..363G} are given in B1950. We transform these coordinates to J2000 ignoring proper motion (for which we have no information at this step) and use a large search radius (15\arcsec) to cross-match the transformed J2000 coordinates with \textit{Gaia} EDR3 (which are provided in epoch 2016). As explained in Section \ref{sec:gaia}, this search radius is large enough to contain most stars at the magnitude range relevant to these polarimetric observations. We then perform a second cross-match within a smaller radius of 10\arcsec comparing the J2000 coordinates of the polarization catalog to the \textit{Gaia} EDR3 coordinates propagated to 2000 (using proper motions and radial velocities as measured in the latter catalog). We found that 10\arcsec allows us to obtain good matches for all but 4 sources in table 1 of \citet{Goodman1990ApJ...359..363G}, 1 source in their table 2. For the remaining sources we assign a the brighest star from \textit{Gaia} within the 15\arcsec radius as the true match and visually inspected the matches. 

Stars in their table 3 appear to have more uncertain coordinates compared to tables 1 and 2. For this reason we increased the cross-match radius to 41\arcsec for these targets. The second cross-match radius also had to be increased compared to the previous tables, to 15 \arcsec. Since the radius has increased, multiple matches can be found for each star. If this was the case we assigned the brightest star as the correct match. This allowed us to find unambiguous matches for all but two stars (80, 88). For these two stars we do not provide a cross-match with \textit{Gaia}. After visual inspection we found that stars 59 and 81 were wrongly matched with a faint star, and for those we used the larger search radius for finding a final match. All coordinates provided for these stars in our Extended Catalog are given in J2000 (after propagation from 2016 with \textit{Gaia} proper motions when available).

We checked for transcription errors in the measurements of \citet{Vrba1976AJ.....81..958V, Moneti1984ApJ...282..508M,Heyer1987ApJ...321..855H,Goodman1990ApJ...359..363G}, by comparing to the entries of $p$ and PA in \citet{Heiles2000AJ....119..923H}. For each star in the aforementioned individual catalogs, we transform its coordinates to J2000, and find the nearest star in the \citet{Heiles2000AJ....119..923H} catalog. We check whether each matched star has $p$ within 0.1\% of the original one and PA within 1 degree (typical uncertainties in the catalog). Most stars pass this criterion. For a few stars the reliable match was the second nearest star - in this case we include these stars in the final catalog. However, stars 23 and 25 from Goodman's table 1 and  stars 12, 52, and 77 from table 3 did not fulfill the criteria for any reliable match. We discarded those stars.

The data in tables 2, 6, 9 and 11 of \citet{Vrba1976AJ.....81..958V} were used in \citet{Goodman1990ApJ...359..363G} and appear to be the same data that are quoted as Goodman et al. (1997) (priv. comm) in \citet{Heiles2000AJ....119..923H}. Stars 70 and 72 of table 2 are given with the same coordinates but have very different polarization properties. We discard these. We transformed the B1950 coordinates to J2000 without the use of proper motion information. We found that the coordinates transformed in this way were accurate enough that the later two-step match with \textit{Gaia} EDR3 (Section \ref{sec:gaia}) provided matches to all these sources.  We found that stars 25, 26, 30, 31, 32, 33, 34, 35, 36, 37 of table 6 do not have any match with \citet{Heiles2000AJ....119..923H} within 30\arcsec. We also note that in this region $-$Lynds 1630$-$ several stars above Dec = 0.1 deg in the \citet{Heiles2000AJ....119..923H} dataset appear in a suspiciously straight line, they have reference to Goodman (1997). However, such a distribution of locations is not seen in the data from \citet{Vrba1976AJ.....81..958V}, indicating a possible transcription error in the coordinates of the \citet{Heiles2000AJ....119..923H} database for these stars. This may be the cause of the lack of a match of the 11 stars of table 6 of \citet{Vrba1976AJ.....81..958V}.

For stars in table 2 of \citet{Heyer1987ApJ...321..855H} we followed the same procedure as for \citet{Vrba1976AJ.....81..958V}, simply transforming their B1950 coordinates to J2000. From \citet{Heyer1987ApJ...321..855H} table 2, we discarded star 53 and from table 3 star 40. Star 56 of table 2 likely has a typographical error in the \citet{Heiles2000AJ....119..923H} catalog (entry 6623) where the $p$ is quoted as 2.3\% whereas it is 2.03\% in the \citet{Heyer1987ApJ...321..855H} table 2. We use the \citet{Heyer1987ApJ...321..855H} values for this star.

We also include data from \citet{Moneti1984ApJ...282..508M}, table 2. No coordinates were provided so we queried Vizier with the HD identifier of each star. We discard star HD283882 of table 2 as its PA is discrepant between that and \citet{Heiles2000AJ....119..923H}. For stars T4, T19 and T29 we were unable to find coordinates and discard them. There were 144 stars in \citet{Moneti1984ApJ...282..508M} table 2 that did not have a match with \citet{Heiles2000AJ....119..923H} within 30\arcsec. We choose to include these data as this is a significant number of measurements.

\subsection{RefID 9: Berdyugin 2001}
We queried Simbad using the star identifiers  provided in this catalog to obtain stellar coordinates. We only kept sources from the NOT program or those unpublished from Haeakala. We excluded sources from  \citep{Markkanen1979,Appenzeller1968,Mathewson1970} as they were already in the \citet{Heiles2000AJ....119..923H} catalog.

\subsection{RefID 18: Alves \& Franco 2006}

The filter used is quoted as 'B'. We assume it is in the Johnsons-Cousins system. Coordinates are in B1950. We convert to J2000 without taking into account proper motions. 

\subsection{RefID 22: Weitenbeck et al 2008}
\label{app:weitenbeck+}
The total uncertainty in Stokes q, u in our final catalog includes a systematic uncertainty of 0.02\% for data in the V,R, and I bands, 0.03\% for the B band and 0.09\% for the UX band as suggested in the text. The systematic uncertainty has been added in quadrature to this given statistical uncertainty. We computed polarization fraction, PA and their uncertainties taking the full error of the Stokes parameters into account. In the end, we also add a systematic $1^\circ$ uncertainty in quadrature to the uncertainty of the PA, as suggested in their text. We have marked as intrinsically polarized the sources that show such polarization from the notes provided on the CDS archive of their paper. In checking for consistency between the published polarization fractions and PA compared to the ones we compute from the Stokes parameters, we found a discrepancy in one measurement for the star HD25427: the PA published for the UX band is given as 130.7$^\circ$, whereas the PA calculated from the Stokes parameters is $139.3^\circ$. As this is the only such discrepancy, we assume that the erroneous published PA of that star is a typographical error.

\subsection{RefID 23: Weitenbeck 2008}
\label{app:weitenbeck}
The published tables in \citet{Weitenbeck2008AcA....58..433W} include measurements of the Stokes parameters, biased polarization fraction and PA. The uncertainties are given for the Stokes parameters (the $e_q$ and $e_u$ are given as equal to each other). For measurements in their table 2 (corresponding to the central wavelength range of the R band, 600-700 nm), we add the systematic uncertainty of 0.02\% to the uncertainty of the Stokes parameters (systematic uncertainty added in quadrature) to obtain the full uncertainty. For data from their table \texttt{pol.dat} we add a systematic uncertainty of 0.02\% for measurements in the R, V and I filters, while it is 0.03\% for the B filter and 0.09\% for the UX filter. We  computed polarization fraction, PA and their uncertainties taking the full error of the Stokes parameters into account.

\subsection{RefID 28: Targon et al. (2011)}

Coordinates were provided in B1950 and we transformed to J2000 ignoring proper motion information. The study targeted fields towards Herbig-Haro objects (HH). We flag each HH object from their table 2 as intrinsically polarized.

\subsection{RefID 29: Santos et al 2011}

The \citet{Santos2011ApJ...728..104S} catalog provides Hipparcos identification numbers. We queried the Hipparcos new reduction catalog \citep{vanLeeuwen2007} to obtain coordinates for each star in J2000. 
Because the coordinates in the Hipparcos catalog are in J1991.25, we transform these coordinates to epoch 2000 using the \texttt{epoch\_prop\_pos} function in the \textit{Gaia} archive ADQL service, inputting the Hipparcos RA, Dec, proper motions and by setting radial velocity = 0 (since Hipparcos does not contain that information). 
More accurate coordinates can be found by querying \textit{Gaia} EDR3, since the latter contains proper motions and radial velocities for a much larger sample of stars than Hipparcos. We thus find the corresponding \textit{Gaia} EDR3 stars by querying the \texttt{gaiaedr3.hipparcos2\_best\_neighbour} catalog, which provides pre-computed cross-matches between EDR3 and Hipparcos. 
There are 146 of the 878 stars that are not contained in this cross-match and so retain the Hipparcos-propagated coordinates. We propagate the \textit{Gaia} EDR3 coordinates of the 732 stars in the cross-match from 2016 to 2000 using proper motion and radial velocity information, when available.

\subsection{RefID 43: Alves et al 2014}
We only include the optical data, not the NIR measurements at wavelengths larger than H band that were in the paper.

\subsection{RefID 48: Serron-Navarrte et al 2016}
We only include data of the V, R, and I bands.

\subsection{RefID 51: Cotton et al 2016}
We divide the uncertainty in the polarization angle by 2, as instructed in the erratum of \citep{Cotton2016MNRAS.455.1607C}. We include the corrected values of HIP2081 from the erratum. Coordinates obtained by querying Simbad. We find a small discrepancy ($\leq 0.2^\circ$) between the quoted position angles from the publication and from the position angles that we compute via the quoted Stokes parameters for some stars. It is unclear what the origin of this discrepancy is, so we use the position angles that we calculate from the Stokes parameters.

\subsection{RefID 54: Wang et al 2017}
We only include data in the optical bands V, i' not the longer wavelength NIR measurements that were also published. We calculated the polarization fraction and angle from the provided Stokes parameters and associated uncertainties. We have checked that this provides consistent polarization fraction with what we obtain by computing the biased polarization fraction from the published values, within rounding errors.

\subsection{RefID 57: Reig et al 2017}
Most data from this paper are for a BeX binary, dominated by circumstellar pol. We only added to our catalog the data for 4 stars in the vicinity of this target (their table A4).

\subsection{RefID 59: Bagnulo et al 2017}

We include data from table A2 of the Bagnulo et al paper. These are synthetic broad-band linear polarization data computed from the full spectral information measured with the FORS2 polarimeter. The authors do not specify the effective wavelengths of the BVRI bands they use. By examining the FORS user manual (table B1) we assume the BVRI refer to the Bessel filters.

\subsection{RefID 63: Neha et al 2018}
\citet{Neha2018MNRAS.476.4442N} provide coordinates for their stars in tables 6, 7. 
However, by cross-matching with \textit{Gaia} EDR3 within 5\arcsec we find that some stars given in the original paper correspond to very faint sources in \textit{Gaia} ($G > 18$ mag). These are in fact mismatches - the actual star observed is brighter, but there was a nearby faint Gaia DR2 in the coordinate search done by \citet{Neha2018MNRAS.476.4442N}. We have confirmed that stars 11, 21, 26 in the original catalog (table 6) are indeed mis-matches (Neha, priv. comm.). 

To recover accurate coordinates, we ran a cross-match of the original catalogs (table 6 and 7) with \textit{Gaia} EDR3 with a search radius of 5\arcsec. In the case when multiple matches were returned for the same star, we selected the brightest star amongst the \textit{Gaia} EDR3 candidates. Stars 11, 21, 26 from table 6 and 192, 193 from table 7 did not return a match. We inspected the area around 20\arcsec from these stars, queried \textit{Gaia} EDR3 and assigned the brightest star with the smallest separation from the original coordinates.
For all stars we obtained the \textit{Gaia} EDR3 coordinates propagated to epoch J2000 from Vizier.

\subsection{RefID 68: Cotton et al. 2019}
Effective wavelength depends on spectral type. Authors give measured effective wavelength for each star, but we only quote an approximate value for each filter for ease of use. Their Table 1 has a typo: RA is actually given in hms units, not dms. Entries in our table have taken the correct units into account.

\subsection{RefID 71: Topasna et al. 2020}
To obtain stellar coordinates (J2000) we queried the WEBDA\footnote{\href{https://webda.physics.muni.cz/webda.html}{https://webda.physics.muni.cz/webda.html}} database and matched the resulting \texttt{No} column values with the given star identifier \texttt{No} of Table 3 in \citet{Topasna2020PASP..132d4301T}.

\subsection{RefID 72: Singh et al. 2020b}
\citet{Singh2020AJ....160..256S} provide \textit{Gaia} DR2 identifiers, so we query \textit{Gaia} DR2, propagating the native coordinates to epoch 2000 (including proper motion information where available). We excluded all sources discussed as intrinsically polarized in their paper (sources with \texttt{S.No.} $\in$ [4, 9, 13, 18, 24, 27, 32, 35, 49, 62, 52, 53, 54, 58, 65, 68, 70, 97, 110]. We use the given Gaia DR2 \texttt{source\_id} as the \texttt{rawstarID} in our catalog.

\subsection{RefID 73: Singh et al. 2020a}
We use the \textit{Gaia} DR2 \texttt{source\_id} provided by \citet{Singh2020AJ....159...99S} as the \texttt{rawstarID} in our catalog. Coordinates in epoch 2000 are obtained following the same process as for \citet{Singh2020AJ....160..256S}. In the case of stars that had no \textit{Gaia} DR2 matches given in their paper, we use their \texttt{S.No.} value as \texttt{rawstarID} in our catalog.

\subsection{RefID 74: Piirola et al. 2020}
\label{app:piirola}
\citet{Piirola2020AA...635A..46P} provide Stokes parameters, as well as the (biased) polarization fraction and polarization angle. Uncertainties are provided for the latter two quantities only. We convert the units of polarization fraction from parts-per-million to fraction (by dividing with $10^6$). We find a slight discrepancy between the published polarization fractions and polarization angles and those we compute from the provided Stokes parameters. This is likely the result of the calibration coefficients, noted in their section 3.1, being applied to the polarization fractions at a later stage without recalculating the Stokes parameters in their work. We choose to rely on the published polarization fractions and angles and compute from those the Stokes parameters and their associated uncertainties. 

\subsection{RefID 78: Singh et al. 2022}
\citet{Singh2022MNRAS.513.4899S} provide \textit{Gaia} DR2 source identifiers for their sources. We queried the \textit{Gaia} archive to obtain DR2 coordinates based on the given identifier. 
We propagate coordinates to epoch 2000 as we did for \citet{Singh2020AJ....160..256S}. We then queried the Gaia DR2-EDR3 cross-match table available on the Gaia archive \texttt{gaiaedr3.dr2\_neighborhood} to obtain Gaia EDR3 identifiers. 
When multiple matches were found for a source, we used the provided column \texttt{magnitude\_difference} to select the match with $\rm min(|magnitude\_difference|)$, ensuring that its value is below 0.5 mag. We did not flag any of their sources as intrinsically polarized, because their assessment was based primarily on discrepancies with the Serkowski law, which could just arise from variations in dust properties.

\bibliography{main}{}
\bibliographystyle{aasjournal}



\end{document}